\def\arxiv{} 

\ifdefined\thesis
    \documentclass[reprint,NumberedRefs]{JASA_blank}
    \def\showfigures{}
\else
\ifdefined\reprint
    \documentclass[reprint,NumberedRefs]{JASA}
    \def\showfigures{}
\else
\ifdefined\reviewers
    \documentclass[preprint,NumberedRefs,TurnOnLineNumbers,trackchanges]{JASA}
    \def\showfigures{}
\else
\ifdefined\submission
    \documentclass[preprint,NumberedRefs,TurnOnLineNumbers]{JASA}
    \def\showfigures{}
\else
\ifdefined\arxiv
    \documentclass[twocolumn,amsmath,amssymb,reprint,aip,floatfix]{revtex4-1}
    \def\showfigures{}
    \usepackage{graphicx}
    \usepackage{caption}
    \usepackage{bm}
    \usepackage{amsfonts}
    \usepackage{natbib}
    \usepackage{latexsym}%
    \usepackage{xcolor}
    \usepackage[bookmarks=false,         
         pdfnewwindow=true,      
         colorlinks=true,    
    final=true
     ]{hyperref}
    \newdimen\reprintcolumnwidth
    \global\reprintcolumnwidth=246pt
    \def\dodoi#1{doi: \href{https://doi.org/#1}{\nolinkurl{#1}}}
    \def\dourl#1{\href{http://#1}{\nolinkurl{#1}}}
\else 
    \documentclass[preprint,NumberedRefs]{JASA}
    \def\appendcaptions{}
\fi
\fi
\fi
\fi

\graphicspath{{.}}

\usepackage{subfigure}

\newcommand{\E}{\mathbb{E}}
\renewcommand{\Vec}[1]{\mathbf{#1}}
\newcommand{\Mat}[1]{\mathbf{#1}}

\renewcommand{\Re}{\mathop{\mathrm{Re}}}
\renewcommand{\Im}{\mathop{\mathrm{Im}}}
\newcommand{\icol}[1]{
  \left(\begin{smallmatrix}#1\end{smallmatrix}\right)%
}
\newcommand*{\inlineref}[1]{%
  \begingroup
    \romannumeral-`\x 
    \setcitestyle{numbers}%
    \cite{#1}%
  \endgroup   
}

\begin{document}

\title[Generative networks with physical priors]{Generative adversarial networks with physical sound field priors}


\author{Xenofon Karakonstantis}
\author{Efren Fernandez-Grande}


\affiliation{Acoustic Technology, Dept. of Electrical \& Photonics Engineering, Technical University of Denmark}




\email{xenoka@dtu.dk}




\begin{abstract}
This paper presents a deep learning-based approach for the spatio-temporal reconstruction of sound fields using Generative Adversarial Networks (GANs). The method utilises a plane wave basis and learns the underlying statistical distributions of pressure in rooms to accurately reconstruct sound fields from a limited number of measurements. The performance of the method is evaluated using two established datasets and compared to state-of-the-art methods. The results show that the model is able to achieve an improved reconstruction performance in terms of accuracy and energy retention, particularly in the high-frequency range and when extrapolating beyond the measurement region. Furthermore, the proposed method can handle a varying number of measurement positions and configurations without sacrificing performance. The results suggest that this approach provides a promising approach to sound field reconstruction using generative models that allow for a physically informed prior to acoustics problems.
\end{abstract}


\maketitle



\section{Introduction}
\noindent
In recent years, the field of machine learning has seen significant progress in its application to inverse problems, fuelled by the increasing accessibility of computational resources and data acquisition methods. While computer vision and imaging have been the primary beneficiaries of these advances, \citep{inverseprob_image} it is becoming increasingly clear that similar approaches can be leveraged to solve inverse problems of physical significance \citep{ardizzoneflows_inverse_problems} or those related to biological and genetic studies. \citep{bioGAN} In the field of acoustics, in particular, machine learning techniques have been utilized to tackle a range of inverse problems, including sound field reconstruction, \citep{sfrecon_inpainting, diegoGPs, XenGAN_internoise} acoustic source localisation, \citep{deepbeamform, ss_vae_localization, GAN_localization} and sound field reproduction and control. \cite{MLacoustics} In this paper, we focus specifically on the application of machine learning for sound field reconstruction.

The acquisition, analysis, and reproduction of sound fields in enclosed spaces such as rooms, vehicle cabins, loudspeaker cabinets, etc. is an important problem in acoustics. It typically involves the reconstruction of acoustic quantities such as the sound pressure, particle velocity, and intensity, all of which provide a detailed description of sound propagation. To define the problem, we assume a spatio-temporal representation of the sound field, typically expressed as a collection of room impulse responses (RIRs) measured or estimated on an aperture which constitutes the spatial domain of interest. For example, this representation is most commonly adopted in sound field synthesis and Ambisonics reproduction, where the listener and the valid sound field are delimited by a space enclosed by a loudspeaker array with far-field sources contributing to the overall reconstructed sound pressure. \cite{ambisonics_wavefield_synthesis}

The reconstruction of sound fields typically relies on the measurement of a selected few positions and the acquisition of the rest of the finite set by either interpolating (interior problem) or extrapolating (exterior problem) the projection of the measured set onto a linear combination of spatial basis functions. These basis functions represent sound propagation in a homogeneous medium where each sound field component is unknown and obtained as part of an optimisation problem. This is commonly brought about as a plane wave expansion, considered an implicit, truncated solution to the homogeneous Helmholtz equation. \cite{samuelCS} A sound field can also be approximated as a finite sum of room modes in a rectangular enclosure \citep{mignot_LFinterpolation} or equivalent spherical sources \citep{antonelloESM} with unknown strength, among other methods. More recent methods propose a representation of spatial convolution with a plane wave kernel for sound field reconstruction. \citep{Manuel_CSC}

Many of the methods mentioned above are considered linear-underdetermined and rank-deficient inverse problems. In other words, the wave basis functions used to represent the sound field are not able to fully span the space of all possible sound fields. This can make the inverse problem underdetermined, where there are multiple solutions that can fit the measurements equally well, or ill-conditioned, where small errors in the measurements can result in large errors in the estimated sound field. To solve this issue, the most common approach is to use regularised least squares methods, such as Tikhonov regularisation or compressive sensing. However, these methods may introduce artefacts in the resulting sound pressure due to under-sampling, implicit assumptions about the distribution of coefficients, or due to the high density of waves above the Schroeder frequency. \cite{aster_parameter_estimation, fjacobsen}

Building upon the capabilities of deep learning algorithms, GANs have emerged as a powerful tool for generating new data instances from underlying distributions, making them ideal for solving problems with difficult-to-observe patterns or features. Despite concerns about their misuse, GANs have demonstrated remarkable data generation capabilities, including in the fields of image and audio synthesis and domain translation. In acoustics-related research, GANs have been used for acoustic metamaterial design, \cite{CaglarGAN, PeterGAN} as well as for room impulse response (RIR) reconstruction using spectro-temporal features. \cite{deep_prior_rir_reconstruction,BW_extension_GAN} In this study, we aim to reconstruct the sound field in a room using a GAN trained on synthetic sound fields, composed of random waves propagating in all three dimensions, and only a few measurements. This approach builds on the stability and optimisation advantages of GANs, which are carried out in the latent space to recover a latent vector that satisfies the measured data, making them ideal for inverse problems.
\section{Method}
\subsection{Plane wave expansion}
\label{subsec:pwexpansion}
The sound pressure $p$ at any position $\Vec{r}_m$ in a room can be approximated by a weighted sum of $N$ plane waves that satisfy the homogeneous Helmholtz equation as \cite{fourieracoustics}
\begin{equation}
    p(\omega , \Vec{r}_m ) \approx \sum_{n = 1}^N x_{n} \text{e}^{j\Vec{k}_{n}^{T}\Vec{r}_m} + n(\omega) \; , 
    \label{eq:planewavesum}
\end{equation}
where $x_{n}$ is the complex plane wave coefficients of the corresponding direction $\Vec{k}_n$ evaluated at position $\Vec{r}_m$ and angular frequency $\omega = k \; c$, the variables $k$ and $c$ are the wavenumber and speed of sound respectively and $n$ is additive Gaussian noise. The plane waves are sampled over a radiation sphere of $N$ discrete directions so that Eq. \ref{eq:planewavesum} converges to the true pressure. \cite{fourieracoustics, plane_waves} We limit the wavenumber spectrum to consider only plane propagating waves – this is a fair assumption considering the acoustic field is evaluated far from sources and boundaries. Hence, Eq. \ref{eq:planewavesum} can be expressed in terms of a matrix formulation so that the measured pressure $\Vec{p} \in \mathbb{C}^{M \times 1} $ at $M$ positions of a spatially extended region $ \mathop{\Omega} \in \mathbb{R}^3$ is
\begin{equation}
    \Vec{p} = \Mat{H} \Vec{x} + \Vec{n} \; , 
    \label{eq:planewaveexpansion}
\end{equation}
where $\Mat{H} \in \mathbb{C}^{M\times N}$ is a dictionary consisted of $N$ plane waves and $\Vec{x} \in \mathbb{C}^{N \times 1}$ is a vector containing the plane wave coefficients.

The problem of recovering the plane wave coefficients is typically ill-posed, rank deficient and typically underdetermined as the measured pressure positions are significantly less than the number of plane waves used to reconstruct the sound field. To obtain a unique and stable solution, additional measurements or regularisation techniques are needed. Regularisation methods involve adding a penalty term to the optimisation problem \cite{aster_parameter_estimation}, which encourages certain properties of the solution, such as smoothness or sparsity. The most commonly used regularisation method is expressed as a regularised minimisation problem, which involves finding the plane wave coefficients that minimise the difference between the measured data and the reconstructed data, while also penalising the weighted values of the estimated model parameters. This can be expressed as
\begin{equation}
    \Vec{\tilde{x}}=\underset{\Vec{x}}{\text{argmin}} \|\Mat{H} \Vec{x} - \Vec{p}\|^2_2, + \lambda \|\Mat{L}\Vec{x}\|_{\mathtt{p}}\; ,
    \label{eq:regularisedLS}
\end{equation}
where $\| \; . \; \|_{\mathtt{p}}$ is the $\ell_\mathtt{p}$ norm ($\mathtt{p} = 1, 2, \cdots , \infty$) and $\Mat{L} \in \mathbb{R}^{N \times N}$ is a weighting matrix commonly selected as an identity matrix or a finite difference operator, to penalise the gradient of the estimated model parameters $\Vec{x}$.

The solution to Eq. \ref{eq:regularisedLS} has a closed-form denoted by the following relation \cite{aster_parameter_estimation}
\begin{equation}
   \Vec{ \tilde{x}} = (\Mat{H}^\mathtt{H} \Mat{H} + \lambda \Mat{L}^\mathtt{T} \Mat{L})^{-1} \Mat{H}^\mathtt{H} \Vec{p}
   \label{eq:LSsolution} \; ,
\end{equation}
where $\Mat{H}^\mathtt{H}$ refers to the hermitian transpose of $\Mat{H}$. It can be assumed that for the various $\ell_\mathtt{p}$ norm regularisation schemes the matrix $\Mat{L}$ is also adapted and weighted accordingly. After finding the coefficients $\Vec{ \tilde{x}}$, they can be used with a new dictionary $\Mat{H}_{*}$ to estimate the sound field at other locations $\Vec{r}_{*}$ bounded by domain $\Omega$. 
\subsection{Generative adversarial networks}
\noindent
Generative Adversarial Networks, or GANs, are a family of deep generative models based on differentiable generator networks. \cite{goodfellow_deeplearning} Other deep generative models include Variational Auto-Encoders (VAEs), \cite{vae} flow based generative models \cite{realnvp} and score-based diffusion models. \cite{denoising_diffusion} GANs consist of two competing networks: a generator network $G$ which generates synthetic data, and a discriminator network $D$ which attempts to classify if observed data are real or synthesised (fake). Hence, GANs are based on a game theoretic scenario in which the generator network must compete against its adversary, the discriminator network. Typically, these networks are involved in a zero-sum game, where each attempts to maximise its own pay-off until convergence, where for a differentiable cost function $J$ we obtain the optimal generator $G^{*}$ as
\begin{equation}
G^{*}= \arg \underset{G}{\min } \max _{D} J(G, D).
\end{equation}
The function $J$ is selected to promote the adversarial game and would commonly be expressed by the relation 
\begin{align}
J(D) &= \mathbb{E}_{{\bf u}\sim f_U({\bf u})} [\log(D({\bf u}))] +\mathbb{E}_{{\bf v}\sim f_V({\bf  v})}[\log(1\!\!-\!D({\bf  v}))], \nonumber \\
J(G) &= -\mathbb{E}_{{\bf v}\sim f_V({\bf  v})}[1 - \log(\!D({\bf  v}))]
\label{eq:GAN}
\end{align}
%
so that the discriminator $D$ is encouraged to characterise the real data $\bf u$ as real and the respective fake data $\bf v$ as fake, while the generator attempts to fool the discriminator into believing that the generated samples with probability density $f_V(\bf v)$ are real. Similarly, the probability density of the real samples is denoted by $f_U(\bf u)$. Although heuristic formulations exist that equate the Adversarial game to maximum likelihood estimation, by minimising the Jensen-Shannon divergence, in practice, the generator simply aims to increase the log-probability that the discriminator misevaluates the generated sample, rather than aiming to decrease the log-probability that the discriminator makes the correct prediction.

As GANs are notoriously unstable during training time, several methods have expanded on the original model by including terms or training schemes that prevent the generator and discriminator from getting stuck in a bad equilibrium. To overcome this issue, various methods have been proposed to improve the stability of GANs, such as the relativistic average GAN (RaGAN). \cite{ralativistic_avg_gan} The RaGAN discriminator estimates the probability that the given real data is more realistic than fake data, on average, instead of comparing them independently. This approach leads to more stable training and better generator performance than a typical GAN. Additionally, it emphasises the difference of the average discriminator output for real and fake data, which helps the model to converge faster. The training objective adopted for training the GANs in this study is given by
%
%
%
%
%
\begin{align}
D_{ra}({\bf u},{\bf v}) &= \operatorname{sigmoid}\left(D({\bf u}) - \mathbb{E}_{{\bf v} \sim f_V({\bf v})}[D({\bf v})]\right) \\
D_{ra}({\bf v},{\bf u}) &= \operatorname{sigmoid}\left(D({\bf v}) - \mathbb{E}_{{\bf u} \sim f_U({\bf u})}[D({\bf u})]\right) \\
J(D) &= \mathbb{E}_{{\bf u} \sim f_U({\bf u})}\left[\log D_{ra}({\bf u},{\bf v})\right] \nonumber \\
&- \mathbb{E}_{{\bf v} \sim f_V({\bf v})}\left[\log(1-D_{ra}({\bf v},{\bf u}))\right] \\
J(G) &= \mathbb{E}_{{\bf u} \sim f_U({\bf u})}\left[\log(1-D_{ra}({\bf u},{\bf v}))\right] \nonumber \\
&- \mathbb{E}_{{\bf v} \sim f_V({\bf v})}\left[\log D_{ra}({\bf v},{\bf u})\right]
\label{eq:relativistic_avg_gan}
\end{align}
where $D_{ra}({\bf v},{\bf u})$ and $D_{ra}({\bf u},{\bf v})$ are the relativistic discriminator scores for a real and fake sample, respectively. The sigmoid function is used as the respective losses assume cross-entropy. To further improve the stability of the discriminator, we employ spectral normalisation\cite{spectralNorm} so it satisfies the Lipschitz constraint \textsc{SN}$(W_D) =1$ where $W_D$ are the weights of the discriminator network. 

As such, the GAN generator $G$ is employed to synthesise the plane wave coefficients and their relative distributions. This is done by assuming a pseudo-complex valued generator network $G$ which takes as an input a latent variable $\Vec{z} \in \mathbb{R}^{l}$, drawn from a well-defined distribution (e.g. ${\cal N}(\Vec{0}, \Mat{I})$) so that the generator network learns to map the noise $\Vec{z}$ to a probability distribution corresponding to plane wave coefficients.
\newcommand{\captionone}{(Color online) The GAN training process. The generator network is trained to synthesise plane wave coefficients, which are then projected onto a plane wave basis to obtain a synthetic sound field. This sound field is evaluated by the discriminator network, which in turn is trained on simulated random wave fields (denoted by ´Real Sound Fields'). By training the generator to produce sound fields that are indistinguishable from the simulated wave fields, one can create realistic sound fields.}
\ifdefined\showfigures
\begin{figure}[!t]
    \includegraphics[width=\reprintcolumnwidth]{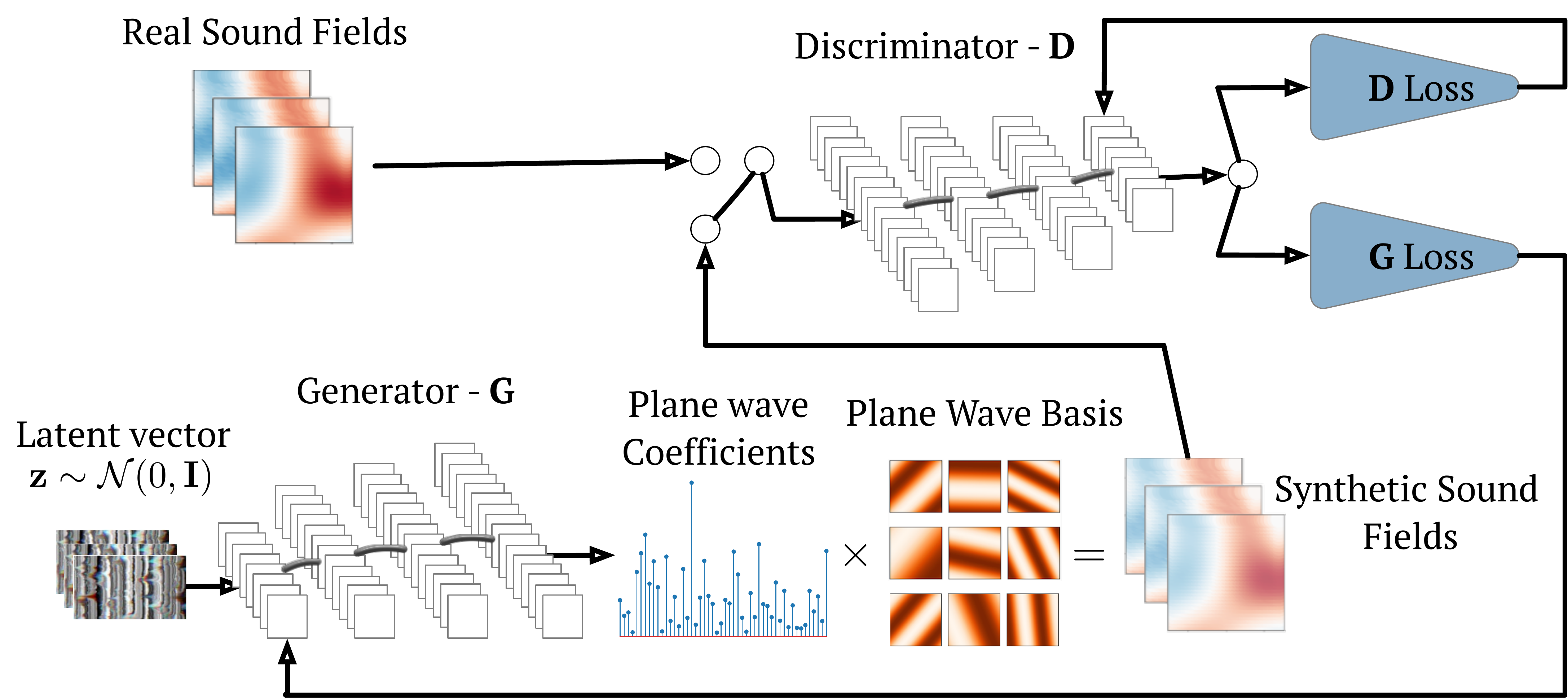}
    \caption{\captionone}
    \label{fig:GANstructure}
\end{figure}
\fi
\subsection{Reconstruction of sound fields using deep generative networks}
Compressive sensing refers to recovering or reconstructing a signal from scarce measurements with solutions to an underdetermined problem \cite{csense} that is bounded by the Nyquist-Shannon sampling criteria. This is carried out by exploiting the sparsity of a signal, leading to the minimisation of Eq. \eqref{eq:regularisedLS} by selecting $\mathtt{p} = 1$ (e.g. $\| x \|_1 \simeq  \| x \|_0$ ). Fundamentally, this corresponds to applying an $\ell_1$ norm penalty and applying regression algorithms such as LASSO \cite{lasso}, as the $l_0$ norm equivalent is NP-hard. 

In the context of sound field reconstruction using a dictionary of plane waves, it is important to consider the limitations of existing theoretical guarantees provided by sparse representation techniques. These guarantees are often derived through incoherence analysis, such as mutual coherence and the Restricted Isometry Property (RIP), \cite{elad2010sparse,yang2018sparsedoa} which are applicable primarily in cases involving incoherent dictionaries. However, in the case of sound field reconstruction, where the atoms of the dictionary $\Mat{H}$ are fairly coherent, these guarantees may not be directly applicable.

Nevertheless, it is worth noting that the lack of theoretical guarantees does not imply that satisfactory performance cannot be achieved in sound field reconstruction. Sparse signal recovery techniques are typically evaluated based on the reconstruction error of the sparse signal $\Vec{x}$, where even a slight error in the support (the set of active components) can lead to a significant estimation error. However, in the specific case of sound field estimation problems, the estimation error is actually measured by the error in the reconstructed acoustic field, i.e., the pressure $\Vec{p}$. As a result, a small estimation error in the support can still be considered acceptable.

To ensure the desired properties of deep generative models for compressive sensing algorithms, certain distribution properties of the weight matrix $\Mat{W}$ in the neural network are crucial. Such a matrix $\Mat{W} \in \mathbb{R}^{N_w \times K_w}$ satisfies the (normalised) Weight Distribution Condition (WDC) with parameter $\epsilon$ if, for all nonzero vectors $\Vec{a}, \Vec{b} \in \mathbb{R}^{K_{w}}$, the following inequality holds \citep{scarlett2022theoretical}
\begin{equation}
    \left\|\frac{1}{N_w} \Mat{W}_{+, \Vec{a}}^\mathtt{T} \Mat{W}_{+, \Vec{b}}-Q_{\Vec{a}, \Vec{b}}\right\| \leq \epsilon,  \label{eq:WDC}
\end{equation}
where $Q_{x,y} = \frac{1}{N_w}\mathbb{E}[\Mat{W}_{+,\Vec{a}}^\mathtt{T} \Mat{W}_{+,\Vec{b}}]$ is the expected value of the product of the positive entries of $\Mat{W}$ associated with vectors $\Vec{a}$ and $\Vec{b}$.

By satisfying the WDC, it is ensured that the weight matrix $\Mat{W}$ possesses certain distribution properties that contribute to the desired properties for deep compressive sensing algorithms. It's important to note that the WDC complements the Range Restricted Isometry Condition (RRIC) and together they form properties analogous to the RIP condition in the compressive sensing literature. \citep{scarlett2022theoretical}

A matrix $\Mat{H} \in \mathbb{R}^{M \times N}$ satisfies the RRIC with respect to a generator network $G$ and constant $\epsilon$ if, for any vectors $\mathbf{z}_1, \mathbf{z}_2, \mathbf{z}_3, \mathbf{z}_4 \in \mathbb{R}^l$, the following inequality holds
\begin{align}
& \mid\left\langle\Mat{H}\left(G\left(\mathbf{z}_1\right)-G\left(\mathbf{z}_2\right)\right), \Mat{H}\left(G\left(\mathbf{z}_3\right)-G\left(\mathbf{z}_4\right)\right)\right\rangle \nonumber \\
& \quad-\left\langle G\left(\mathbf{z}_1\right)-G\left(\mathbf{z}_2\right), G\left(\mathbf{z}_3\right)-G\left(\mathbf{z}_4\right)\right\rangle \mid \nonumber \\
& \quad \leq \epsilon\left\|G\left(\mathbf{z}_1\right)-G\left(\mathbf{z}_2\right)\right\|_2\left\|G\left(\mathbf{z}_3\right)-G\left(\mathbf{z}_4\right)\right\|_2. 
\label{eq:RRIC}
\end{align}
In simple terms, the RRIC condition assesses the behaviour of the matrix $\Mat{H}$ in relation to the range of the mapping function $G\left(\cdot \right)$. It states that if we take four vectors, $\Vec{z}_1, \Vec{z}_2, \Vec{z}_3,$ and $\Vec{z}_4$, and apply the mapping  $G\left(\cdot \right)$ to them, the difference between the inner product of $\Mat{H}$ acting on the differences of the mapped vectors and the inner product of the original differences should be small compared to the norms of the mapped differences. In other words, the RRIC condition ensures that the matrix $\Mat{H}$ acts similarly to an isometry when operating on pairs of differences of signals obtained by applying the mapping $G\left(\cdot \right)$. An isometry preserves distances between vectors, and the RRIC condition guarantees that the distances are not significantly distorted when applying $\Mat{H}$ to the differences of mapped signals.

In our study, the primary objective is to recover the complex plane wave coefficients $\hat{\Vec{x}} = G(\Vec{z}^*)$ to decompose the measured pressure $\hat{\Vec{p}} = \Mat{H} \hat{\Vec{x}}$ (similar to Eq. \ref{eq:planewaveexpansion}). We aim to find the latent representation of the optimal coefficients $\Vec{z}^* \in \mathbb{R}^{l}$ through empirical risk minimization. This involves minimizing the misfit between the estimated and measured pressures, achieved via the injective mapping $G: \mathbb{R}^l \rightarrow \mathbb{C}^N$.

To accomplish this, we consider a $d$-layered generative network modeled by $G (\Vec{z}) = \sigma(\Mat{W}_d \cdots \sigma(\Mat{W}_2 \, \sigma(\Mat{W}_1 \Vec{z}) \, ) \cdots )$. Here, $\sigma(\cdot)$ represents a non-linear activation function, and $ \Mat{W}i \in \mathbb{R}^{N_i \times N{i-1}}$ for $i \in {1,\cdots,d}$ are the weights of each layer. This network maps a latent variable following a known distribution $\Vec{z} \sim \mathcal{N}(0,\Vec{I})$ to an unknown distribution, which corresponds to the plane wave coefficients propagating in the measured sound field. This process can be described by the equation
\begin{equation}
\Vec{\tilde{z}}=\underset{\Vec{z}}{\text{argmin}} \|\Mat{H} G(\Vec{z}) - \Vec{p}\|^2_2, + \lambda \|\Vec{z}\|_2.
\label{eq:CSGAN}
\end{equation}
The $\ell_2$ penalty on the latent variable $\Vec{z}$ enforces a Gaussian constraint on its prior distribution. By assuming that the true plane wave coefficients $\Vec{x}^*$ lie near the range of distributions spanned by the generative model $G(\Vec{z}^*)$, one can stochastically extrapolate the sound field in the measured room. By ensuring sufficient expansivity, which refers to an increase in the number of nodes or filters from one layer to another, along with the aforementioned conditions of Eqs. \eqref{eq:WDC} and \eqref{eq:RRIC}, it can be guaranteed that Eq. \eqref{eq:CSGAN} does not possess any undesirable local minima beyond the solutions represented by $\mathbf{z}$ and its negative multiple. To meet this criterion, it is typical for the generator network $G$ to employ Gaussian i.i.d. weights, while the matrix $\Mat{H}$ should consist of Gaussian i.i.d. entries with a substantial number of columns. Although a previous study \citep{HandVoroninski} has observed certain trained networks displaying statistical attributes reminiscent of a Gaussian distribution, it is important to note that modelling $\Mat{H}$ as a random measurement matrix introduces the notion of completely incoherent atoms. However, as mentioned earlier, this incoherence assumption does not hold in the context of acoustics problems. \citep{yang2018sparsedoa}

Shamshad et al. proposed a variant of Eq. \eqref{eq:CSGAN} that allows for a generalisation of the problem, by extending the span of the generator network. \cite{deepPtych} This is accomplished by extending the span of $G$ by a vector with unknown parameters obtained as part of a two-step optimisation process. This can be described by
\begin{equation}
\Vec{\tilde{\Vec{q}}}=\underset{\Vec{q}}{\text{argmin}} \|\Mat{H} \Vec{q} - \Vec{p}\|^2_2, + \lambda_{\Vec{q}} \| \Vec{q} - \hat{\Vec{q}} \|_2 \; ,
\label{eq:deepPtych}
\end{equation}
where $\Vec{q}$ is now the unknown vector of plane wave coefficients to be recovered and $\hat{\Vec{q}} = G(\tilde{\Vec{z}})$ is the vector of complex coefficients obtained in the previous optimisation objective of Eq.\,(\ref{eq:CSGAN}). Now the model is able to deviate from the span of the generator network, and the hyperparameter $\lambda_{\Vec{q}}$ controls how strictly this is applied. This means that even if the generator $G$ has not seen similar distributions to that of the measured sound field $\Vec{p}$ (e.g. in the training data), the coefficients can adapt to the data by finding the projection $\Vec{q}$ onto the data subspace but constraining it to stay close to the solution found previously by the generator $\hat{\Vec{q}}$. The objective of Eq.\,\eqref{eq:deepPtych} is solved once again with stochastic gradient descent and the auto-differentiation algorithms inherent in most machine learning libraries. Finally, the interpolated or extrapolated pressure is obtained as  
\begin{equation}
    \hat{\Vec{p}}= \Mat{H}_{*} \Vec{\tilde{q}} \; ,
\end{equation}
so that $\Mat{H}_{*}$ is a plane wave transfer matrix projecting the coefficients $\Vec{\tilde{q}}$ to positions $M^{*}$.
\section{Network Training and Inference}
\subsection{GAN architecture} \label{subsec:GAN_architecture}
This work employs a GAN reminiscent of a Deep Convolutional Generative Adversarial Network (DCGAN) architecture, \cite{DCGAN} modified to handle complex-valued data and feature spaces. The complex-valued data are obtained by concatenating the real and imaginary parts channel-wise. The GAN discriminator network evaluates the quality of the sound fields synthesised by the generator network. The generator output, a coefficient distribution, is passed through a transfer matrix $\Mat{H}_{g}$ to obtain the final synthetic sound field (i.e., $\Vec{v} \equiv \Mat{H}_{g} G(\Vec{z})$). The generator network uses Rectified Linear Units (ReLU) as the activation function for all layers, except the last layer, which uses a linear activation function. Additionally, Instance Normalization (IN) is applied in all layers of the generator network. On the other hand, the discriminator network uses Leaky ReLUs (LReLU) with a slope of $\alpha = 0.2$ and spectral normalisation in all convolutional layers, again with the exception of the last, fully connected (FC) layer which uses a linear activation map. The detailed architecture of both networks can be found in Table \ref{tab:GANarchitecture}.

\begin{table}[ht]
\caption{Architecture of GAN used for plane wave expansion, with both the discriminator $D$ (top) and the generator $G$ (bottom).}
\label{tab:GANarchitecture}
\centering

\resizebox{\reprintcolumnwidth}{!}{%
\begin{tabular}{lll}
Layer & Filters & Output Size \\
\hline\hline \textbf{Discriminator} $D$: & & \\
\hline\hline 
Planar Sound Field $\Vec{p}$ & & $21 \times 21 \times 2$ \\
\hline
Conv-1 + Stride 1 + LReLU(0.2) & $4 \times 4 \times 32$ & $21 \times 21 \times 32$ \\
Conv-2 + Stride 2 + LReLU(0.2) & $4 \times 4 \times 32$ & $11 \times 11 \times 32$ \\
Conv-3 + Stride 2 + LReLU(0.2) & $4\times 4 \times 64$ & $6 \times 6 \times 64$ \\
Conv-4 + Stride 2 + LReLU(0.2) & $4 \times 4 \times 128$ & $3 \times 3 \times 128$ \\
Conv-5 + Stride 2 + LReLU(0.2)& $4 \times 4 \times 256$ & $2 \times 2 \times 256$ \\
Conv-6 + Stride 2 + LReLU(0.2)& $4 \times 4 \times 512$ & $1 \times 1 \times 512$ \\
Reshape + FC + LReLU(0.2)& $ 512 \times 100$ & $100$ \\
FC & $100 \times 1$ & 1 \\

\\
\hline\hline \textbf{Generator} $G$: & & \\
\hline\hline
Latent Variable $\bf z$ & &  $128 \times 1$ \\
\hline
ConvTranspose-1 + Stride 1  + IN + ReLU & $4 \times 4 \times 1024$ & $4 \times 4 \times 1024$ \\
ConvTranspose-2 + Stride 2  + IN + ReLU & $4 \times 4 \times 512$ & $8 \times 8 \times 512$ \\
ConvTranspose-3 + Stride 2  + IN + ReLU & $4 \times 4 \times 256$ & $16 \times 16 \times 256$ \\
ConvTranspose-4 + Stride 2  + IN + ReLU & $4 \times 4 \times 128$ & $32 \times 32 \times 128$ \\
ConvTranspose-5 + Stride 2 & $4 \times 4 \times 2$ & $64 \times 64 \times 2$ \\
Reshape & $-$ & $4096 \times 2$
\end{tabular}
}
\end{table}
\subsection{GAN training} \label{subsec:GANtrain}
\noindent
The sound fields used in our training dataset are numerical representations of sound pressure measurements, generated in real-time using a simulation of a truncated sum of plane waves arriving from various directions, with randomised amplitudes and a uniform phase as introduced by Fernandez-Grande et al. \cite{BW_extension_GAN} These arrays have a shape of $[21 \times 21]$, and the measurements were taken at uniformly spaced intervals of 0.05 m, resulting in an aperture of $1 \times 1$ m$^2$.
%

During the training and inference process, the network uses complex-valued computations as the data is separated into its real and imaginary components and concatenated channel-wise, i.e. $ \Vec{p} = \icol{\Re( \Vec{p} )\\ \Im ( \Vec{p} )}$. However, the convolutional kernels used by the network are only capable of processing real-valued data, which leads to an approximation of the true complex-valued data.

The generator network is trained using an adversarial objective for over 40000 iterations, which is carried out over 24 hours on a single GPU. The training dataset includes sound fields that are evenly distributed across different frequencies. The training process is demonstrated in Fig. 1. It is important to note that the GAN is not trying to match any specific pressure measurements during this stage, but rather it is learning the underlying statistical distributions of the pressure.

To apply the learned coefficients to each batch of sampled dictionaries, we sample 4096 plane wave directions uniformly from the surface of a sphere with a radius of $k$, which corresponds to the wavenumber. The number of plane waves used corresponds to the vectorized output of the generator network (e.g., $N = 64 \times 64 = 4096$) and thus the number of coefficients. The number of waves necessary to represent a sound field, as approximated by Vekua theory \citep{vekua2}, is given by $ N \simeq \left(\left\lceil ka_{\Omega}\right\rceil+1\right)^2$, \citep{vekua1} where $a_{\Omega}$ is the radius of the reconstruction domain $\Omega$ and $\lceil \cdot \rceil$ denotes the ceiling integer round-off function. For example, with a radius of $a_{\Omega} = 0.7$ m and an upper frequency limit of $f = 4$ kHz, the estimated number of waves is approximately 2735.

Regarding the generator network architecture, each 2D feature map dimension is upsampled by a factor of 2 in every layer, resulting in a total dimensionality multiplication factor of 4. The output sizes of layers $1, 2, \dots, d$ correspond to $16, 64, 256, 1024$, and $4096$, respectively. Notably, the final output size (4096) exceeds the representation limits defined by the aforementioned equation for $N$. It bears mentioning that any feed-forward network could be used instead of the generator network, as the network does not need any specific spatial information or translation invariance. However, we use a convolutional network in this case, as it allows us to utilise parameter sharing and thus reduces memory requirements. Furthermore, it is likely that neighbouring features within the convolutional neural network would exhibit common statistical attributes, reflecting a physical interpretation where neighbouring waves in the sound field also share common statistical properties. The code has been made available in Ref \inlineref{PWGANcode}.

For evaluating the generative capacity of the GAN, a method for comparing distributions is necessary. To this end, we use the Maximum Mean Discrepancy (MMD) due to its strong theoretical properties \citep{MMD} and ease of implementation. The MMD calculates the distance between two distributions, $P(X)$ and $Q(Y)$, by measuring the squared distance between the means of their embeddings. Mathematically, the MMD between distributions $P$ and $Q$ is computed as
\begin{equation}
\operatorname{MMD}(P, Q)=\left\|E_{X \sim P}[\phi(X)]-E_{Y \sim . Q}[\phi(Y)]\right\|, \label{eq:MMD}
\end{equation}
where $X$ and $Y$ are samples drawn from distributions $P$ and $Q$, respectively, and $\phi$ represents the embedding function or kernel.

It is important to note that the embeddings produced by the function $\phi$ must reside in a Reproducing Kernel Hilbert Space (RKHS). The RKHS provides the necessary mathematical structure for capturing and comparing the distributions effectively. In this case, the kernel to evaluate the distributions of the generated data against the training data is the radial basis function (i.e., $k(x, y) = \exp\left(-\gamma |x-y|^2\right)$), where $\gamma$ is obtained by inverting the square of the median estimate of the piecewise differences in the training data samples. By computing the MMD between distributions, we can quantify the dissimilarity or discrepancy between them. This is particularly useful in various tasks such as distribution alignment, generative modeling, or measuring the discrepancy between real and generated data distributions.

Finally, Both the discriminator and generator were trained with a learning rate of $\eta_{D} = \eta_{G} = 0.0002$ and the minibatch size for both networks was set to $32$. Figure 2 shows examples of random sound fields generated by the generator $G$ for a fixed latent variable $\Vec{z}$ and $f = 500$ Hz. The generated sound fields begin to resemble the training data quite early on during training, as established when examining the MMD between training and generated data of Fig. 2.

\newcommand{\captiontwo}{(Color online) Examples of the real components of the generated sound fields (top) for 100, 1000, 10000 and 40000 iterations of training. The latent variable is the same for each example and the frequency displayed is $f = 500$ Hz and the MMD between training samples and generated samples during the training phase (bottom).}
\ifdefined\showfigures
\begin{figure}[!t]
    \includegraphics[width=\reprintcolumnwidth]{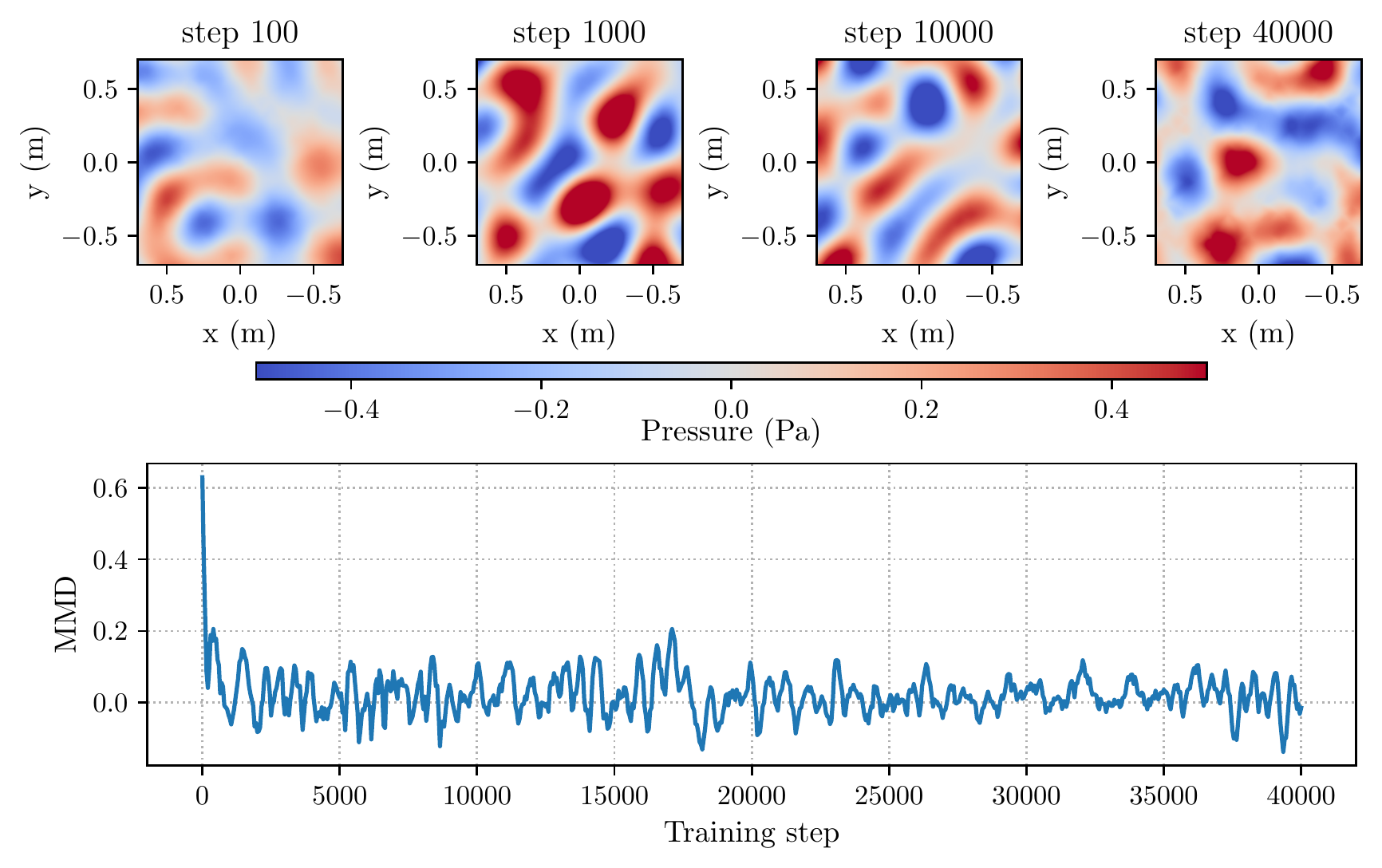}
    \caption{\captiontwo}
    \label{fig:fig2}
\end{figure}
\fi
\subsection{Experimental validation}
The two datasets used in this work for validating the proposed method are consisted of room impulse responses measured in two different rooms and configurations. Both dataset configurations can be seen in Fig. 3.
\newcommand{\captionthree}{(Color online) The measurement configurations of the validation datasets, (a) DTU dataset\cite{DTUdata} and (b) MeshRIR dataset\cite{MeshRIRdata}}
\ifdefined\showfigures
\begin{figure}
    \centering
    \begin{minipage}[b]{0.44\columnwidth}
        \centering
        \includegraphics[width=\textwidth]{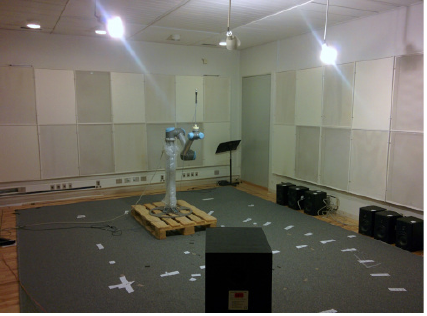}
        \caption*{(a)} 
    \end{minipage}
    \hfill
    \begin{minipage}[b]{0.47\columnwidth}
        \centering
        \includegraphics[width=\textwidth]{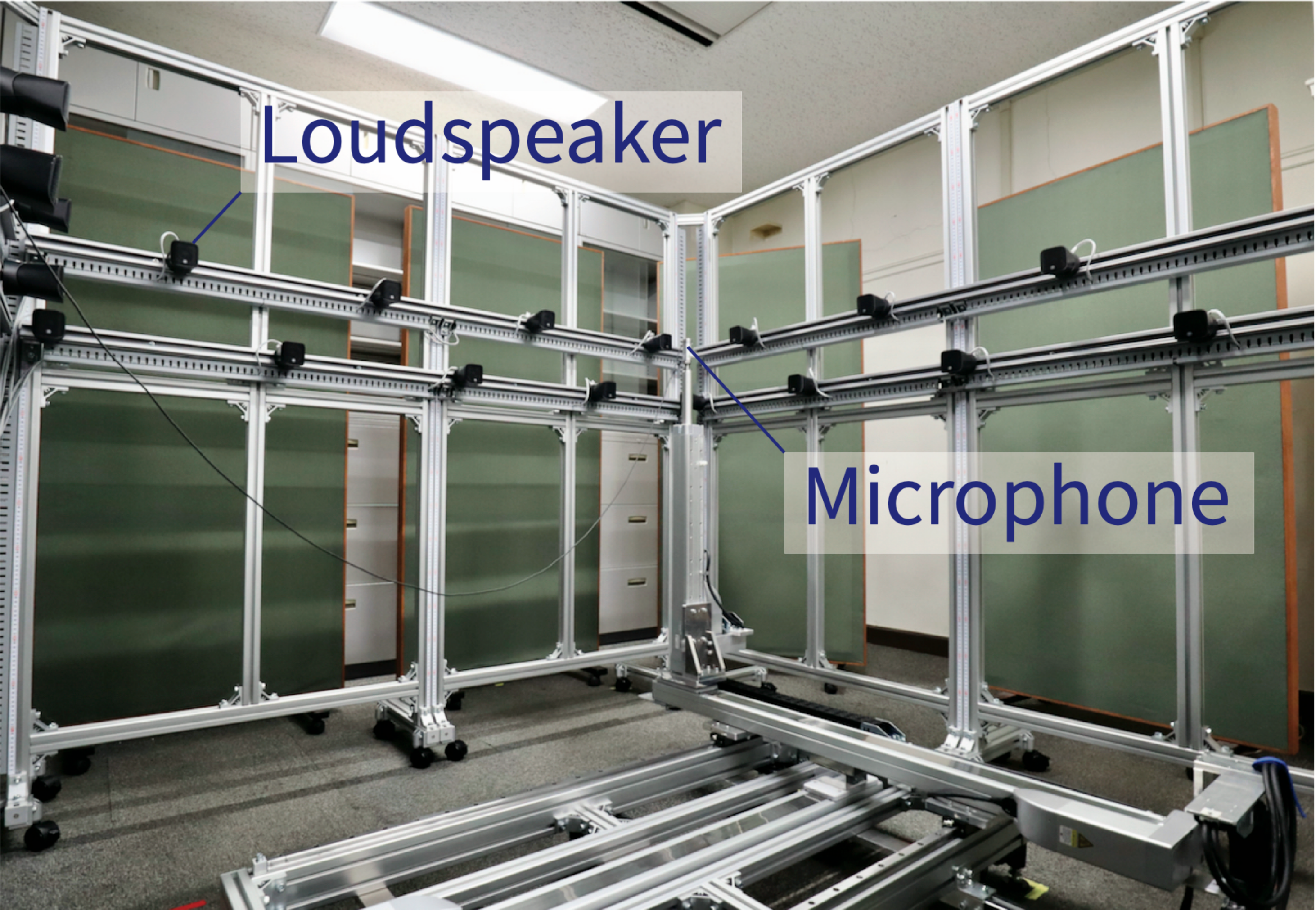}
        \caption*{(b)} 
    \end{minipage}
    \captionsetup{list=no} 
    \caption{\captionthree}
    \label{fig:dataset_photos}
\end{figure}
\fi
\begin{table}[!t]
\renewcommand{\arraystretch}{1.3}
\caption{Size and reverberation times  octave bands from the studied rooms}
\label{tab:room_char}
\centering
\resizebox{\reprintcolumnwidth}{!}{%
\begin{tabular}{|c|c|c|}
\hline
Room & Dimensions & $T_{30}$(s) (125, 250, 500, 1k, 2k, 4k Hz)\\
\hline
DTU & $7.5\times 4.74 \times 2.8$ & 0.5, 0.4, 0.4, 0.4, 0.4, 0.4\\
MeshRIR & $7.0\times 6.4 \times 2.7$ & 0.22, 0.31, 0.35, 0.28, 0.24, 0.2 \\
\hline
\end{tabular}
}
\end{table}

The first dataset, hereby referenced to as DTU dataset, consists of RIRs measured in a listening room that complies with the IEC standard 268-13, as described in Table \ref{tab:room_char}. The room dimensions and reverberation times in octave bands for this dataset can be found in the table. This dataset was obtained from previously published sources\cite{samuelCS} and which is publicly available.\cite{DTUdata} The Data are captured at a 44100Hz sampling frequency and for the purpose of computational efficiency, the RIRs were subsequently resampled to a lower rate of 8000Hz. The DTU dataset was divided into two subsets, a ``fitting set" and a ``reconstruction set", as depicted in Fig. 4. The fitting set consists of 97 measurements taken on a sphere with a radius of 0.5 m, as well as an additional 5 measurements taken within the sphere, known as anchor points. The reconstruction set, on the other hand, includes 703 measurements taken on a circular plane with a radius of 0.7 m located on the equator of the sphere.
\newcommand{\captionfour}{(Color online) DTU dataset layout used for GAN validation}
\ifdefined\showfigures
\begin{figure}[htb!]
    \centering
    \includegraphics[width = \reprintcolumnwidth]{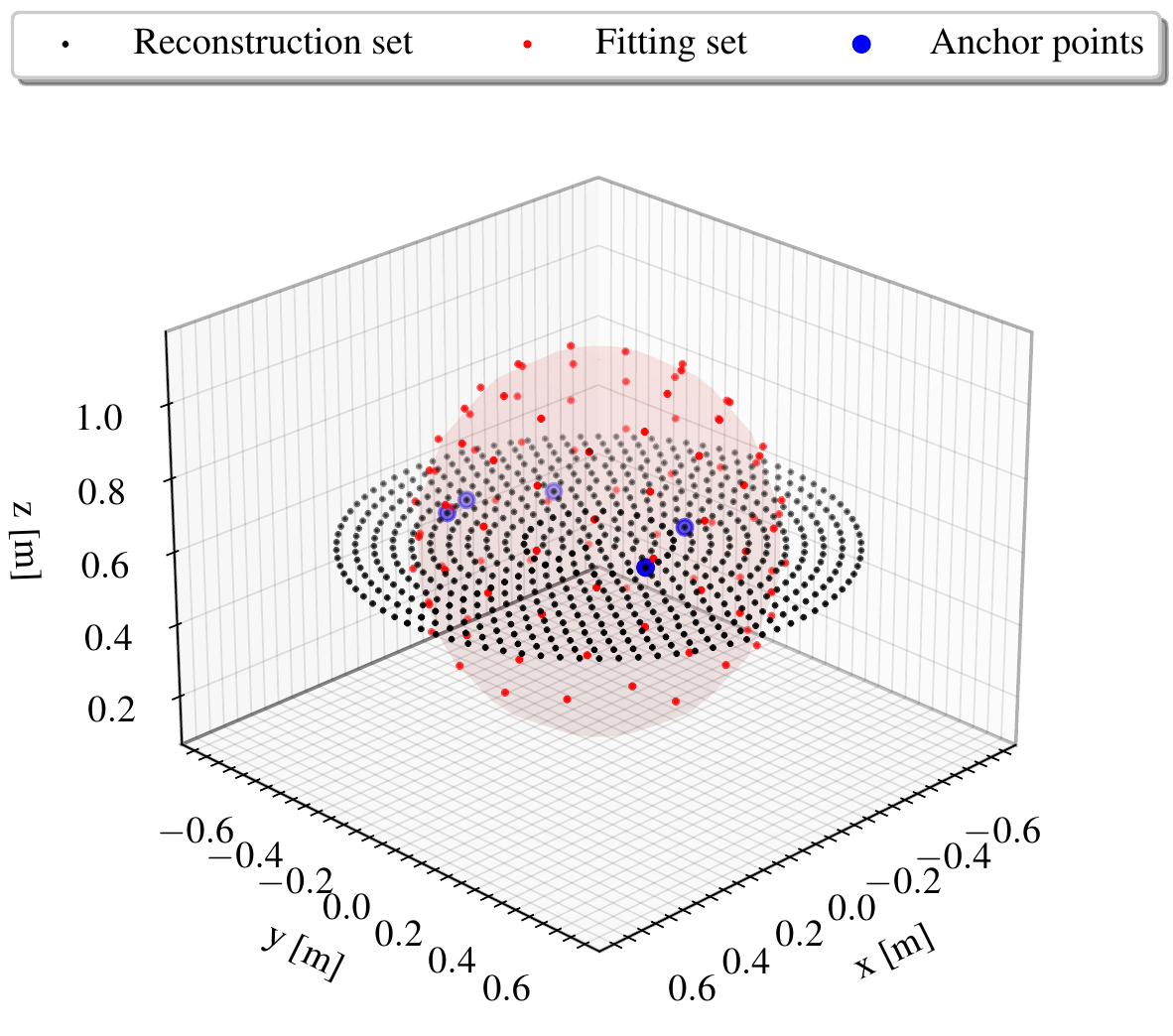}
    \caption{\captionfour}
    \label{fig:DTU_data_layout}
\end{figure}
\fi

The second dataset, MeshRIR, is made publicly available by Koyama et al.\cite{MeshRIRdata} and consists of 441 RIRs for 32 different source configurations captured in a highly damped room with several acoustic absorption panels, as described in Table \ref{tab:room_char}. The reverberation times in octave bands for this dataset can be found in the table. The RIRs were measured inside a 2D square region of 21$\times$21 measurement positions (with intervals of 0.05 m), corresponding to a 1 m $\times$1 m square, using a Cartesian robot. The data were captured at a sampling rate of, 48000Hz and, like the DTU dataset, was also resampled to a lower rate of 8000Hz for computational reasons. Additionally, the MeshRIR dataset was decimated with respect to the number of microphones to evaluate the pressure interpolation capabilities of the proposed GAN-based generator. Therefore, from the complete set of 441 measurements, we decimate at random to obtain datasets with the number of measurements varying between 20 and 245 positions for experimental validation, while using the same source position. The layout of the complete dataset can be seen in Fig. 3.
\newcommand{\captionfive}{(Color online) MeshRIR dataset layout used for GAN validation}
\ifdefined\showfigures
\begin{figure}[htb!]
    \centering
    \includegraphics[width = 0.8\reprintcolumnwidth]{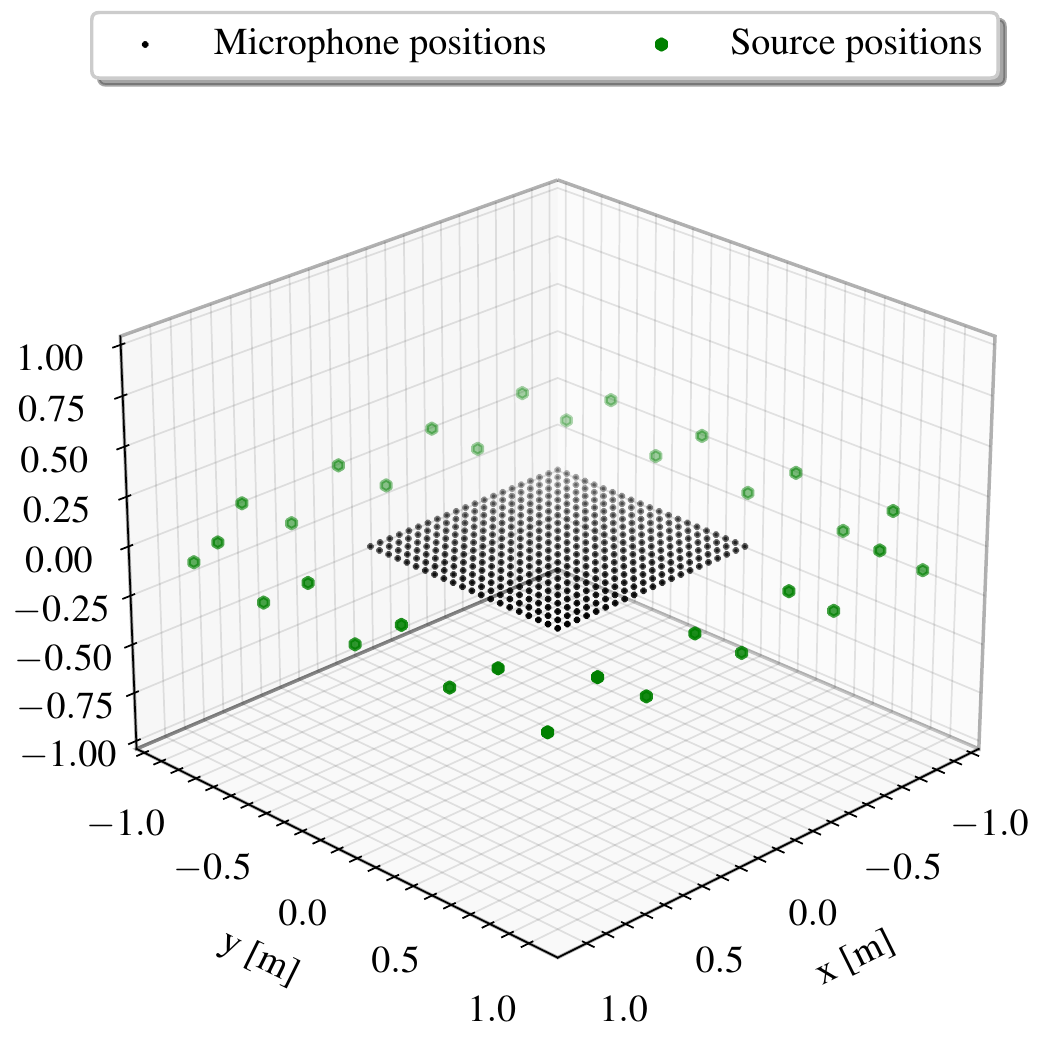}
    \caption{\captionfive}
    \label{fig:MeshRIR_data_layout}
\end{figure}
\fi
%
\subsection{Sound field prediction}\label{subsec:sf_prediction}
\noindent
The hyperparameters for optimizing the Plane Wave GAN (PWGAN) were determined using a grid search with the objective of minimizing the mean square error between observed and reconstructed signals. The optimization process involved sensing matrices with $N=4096$ plane waves. These waves were used to fit the data and project it onto unmeasured planes. 

The optimization process consists of two stages. Initially, 2000 steps of Eq. \eqref{eq:CSGAN} are executed, followed by 2000 steps of Eq. \eqref{eq:deepPtych} once an optimal latent variable $\Vec{z}$ is recovered. During the first stage, a penalty is enforced on the latent variable $\Vec{z}$ weighted by $\lambda_{\Vec{z}} = 0.001$. In the second stage, the penalty is weighted by $\lambda_{\Vec{q}} = 0.005$. The learning rate for the gradient-based algorithm is set to $\eta_{\Vec{q}} = \eta_{{\Vec{z}}} = 0.001$ for both stages.

To evaluate spatio-temporal reconstruction performance in the frequency domain, two commonly used metrics were adopted. The Normalised Mean Square Error (NMSE) measures the discrepancy between each estimated pressure value and scaled by the pressure variance, while the Spatial Similarity (SS), otherwise known as the Modal Assurance Criterion (MAC), describes the similarity on a scale of $0$ to $1$, with $1$ being identical and $0$ fully dissimilar. The aforementioned metrics can be described by the following relations,
\begin{align}
 \textsc{NMSE} &= 10 \log_{10}\left(\frac{\lvert| \Vec{p} - \hat{\Vec{p}} |\rvert^2 }{\lvert| \Vec{p} |\rvert^2 }\right) \; ,
 \label{eq:nmse}
 \\          
 \textsc{SS} &= \frac{\lvert \Vec{p} \hat{\Vec{p}} \rvert^2}{(\Vec{p}^{\mathtt{H}} \hat{\Vec{p}}) (\hat{\Vec{p}} ^{\mathtt{H}} \Vec{p})} \; , 
 \label{eq:ss}
\end{align}
where $\Vec{p}$ and $\hat{\Vec{p}}$ refer to the true and estimated pressures respectively, $\lvert| .|\rvert$ the Euclidean norm, and $\lvert.\rvert$ the absolute value.
For evaluation of the reconstructed RIRs (in time domain) we use Pearson's correlation coefficient defined as
\begin{equation}
    \rho(\mathbf{r}, t) = \frac{\E[p(\mathbf{r}, t)\hat{p}(\mathbf{r}, t)]-\E[p(\mathbf{r}, t)]\E[\hat{p}(\mathbf{r}, t)]}{\sqrt{\E[p^2(\mathbf{r}, t)]\E[p(\mathbf{r}, t)]^2}\sqrt{\E[\hat{p}^2(\mathbf{ r}, t)] - \E[\hat{p}(\mathbf{r}, t)]^2}},
\end{equation}
between any true (reference) $p(\mathbf{r}, t)$ and reconstructed $\hat{p}(\mathbf{r}, t)$ RIR on the reconstruction plane. 

In order to perform a comparison, the algorithm of regularised least-squares (Tikhonov) regression was selected as a baseline for the DTU dataset. For the MeshRIR dataset, two neural networks were used as a baseline, a U-net Auto-Encoder (AE) as proposed by Llu{\'\i}s et al.\cite{sfrecon_inpainting} and a Variational Autoencoder (VAE) from which we perform the latent space optimisation of Eq.\, \eqref{eq:CSGAN} once it is fully trained. Both networks are trained with the ``online'' random wave method proposed in this paper, for reasons of fair comparison. The least-squares regression algorithm assumes that both the coefficients and residuals are i.i.d. and normally distributed. The regularisation parameters for the method were determined by the variance of additive noise and subsequently confirmed through 10-fold cross-validation. We examine the temporal correlation of the reconstructions with respect to the undisclosed to the GAN measurements. 

For the neural networks, the sound field magnitude $\lvert \bf p \rvert$ was reconstructed over the reference plane as depicted in Fig. 5, using the same dimensions prescribed for the GAN training data as outlined in Subsection \ref{subsec:GANtrain} for a varying number of frequencies and fixed microphone positions and vice versa. The reasoning behind selecting the magnitude as opposed to the full pressure is that the  Llu{\'\i}s et al.\cite{sfrecon_inpainting} Auto-Encoder is designed to reconstruct the magnitude, since it uses a strictly positive (sigmoid) activation function as a final activation. It should be noted that the U-net Auto-Encoder was trained by varying both the number and positions of the microphones where the random wave fields were evaluated. Between 20 and 220 microphones were used during training. More details regarding the aforementioned networks can be seen in \ref{Appendix}.

\section{Results}
\noindent
In this section, we present the results of the proposed PWGAN method for sound field reconstruction. The results are evaluated using two datasets: the DTU dataset and the MeshRIR dataset. 

\subsection{Spherical array sound field and RIR reconstruction}
\noindent
For the DTU dataset, the reconstructed room impulse responses (RIRs) are evaluated as a point within the interpolated reconstruction plane (i.e., within the spherical array) and another in the extrapolated reconstruction plane (i.e., outside the sphere), as well as the spatial distribution of the pressure over the whole reconstruction plane.
Figure 6 shows the correlation coefficient between reconstructed and true RIRs as a function of distance from the spherical array centre, averaged over the responses which fall roughly into the same distance (e.g. intervals of 0.05 m). Here, the difference is significant between both examined methods, with the PWGAN demonstrating better performance. As expected, the performance of both algorithms increases closer to the centre of the array and closer to where the microphones are placed ($\lvert r \rvert = 0.5$ m).
\newcommand{\captionsix}{(Color online) Correlation as a function of distance from the spherical array centre. The vertical dashed line denotes the location of the spherical array.}
\ifdefined\showfigures
\begin{figure}[h!]
    \centering
    \includegraphics[trim={0mm 0 0mm 0},clip, width = \reprintcolumnwidth]{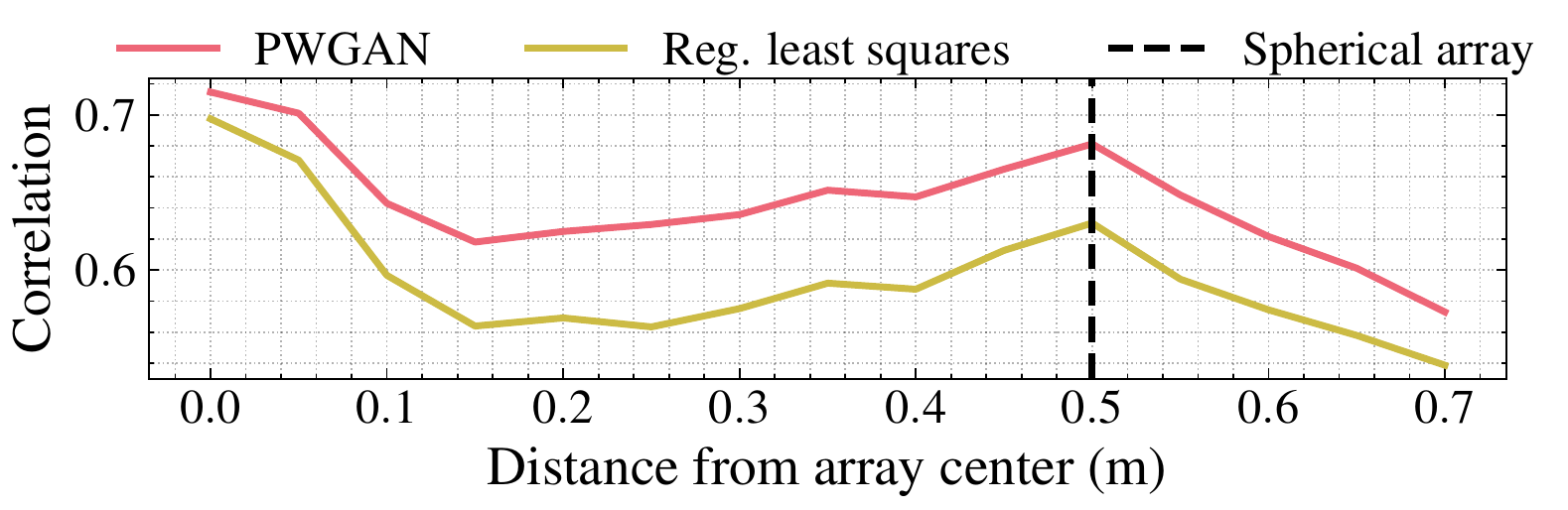}
    \caption{\captionsix}
    \label{fig:corr}
\end{figure}
\fi

Figure 7 displays the interpolated RIRs and their respective frequency response functions (FRFs) at distance $\lvert r \rvert = 0.05$ m from the centre of the spherical array, so that the RIRs have been normalised to a maximum amplitude of 1 and the FRFs to a reference of 1 dB for ease of comparison. At first glance, it can be seen that both methods can reconstruct the RIRs accurately, with the PWGAN reconstruction being slightly more accurate (0.74 correlation as opposed to 0.72 correlation). Additionally, the RIRs of Fig. \ref{fig:InterpRIRFRFcombined} display less pre-ringing in onset from the PWGAN reconstruction in comparison to Reg. least squares reconstruction. The FRFs also reveal that the PWGAN retains more energy at high frequencies (2-4 kHz) than the respective Reg. least squares FRF. However, it can also be seen that the PWGAN displays slightly more erroneous behaviour at low frequencies (below 500 Hz). 
\newcommand{\captionseven}{(Color online) Interpolated RIRs (a) and FRFs (b) for Reg. least squares (yellow) and PWGAN (red) with the respective correlation coefficients of 0.72 and 0.74 at $\lvert r \rvert$ = 0.05 m from the array centre.}
\ifdefined\showfigures
\begin{figure}[!ht]
\centering
\begin{subfigure}
\centering
\includegraphics[trim={0mm 1mm 1.5mm 0},clip, width=\linewidth]{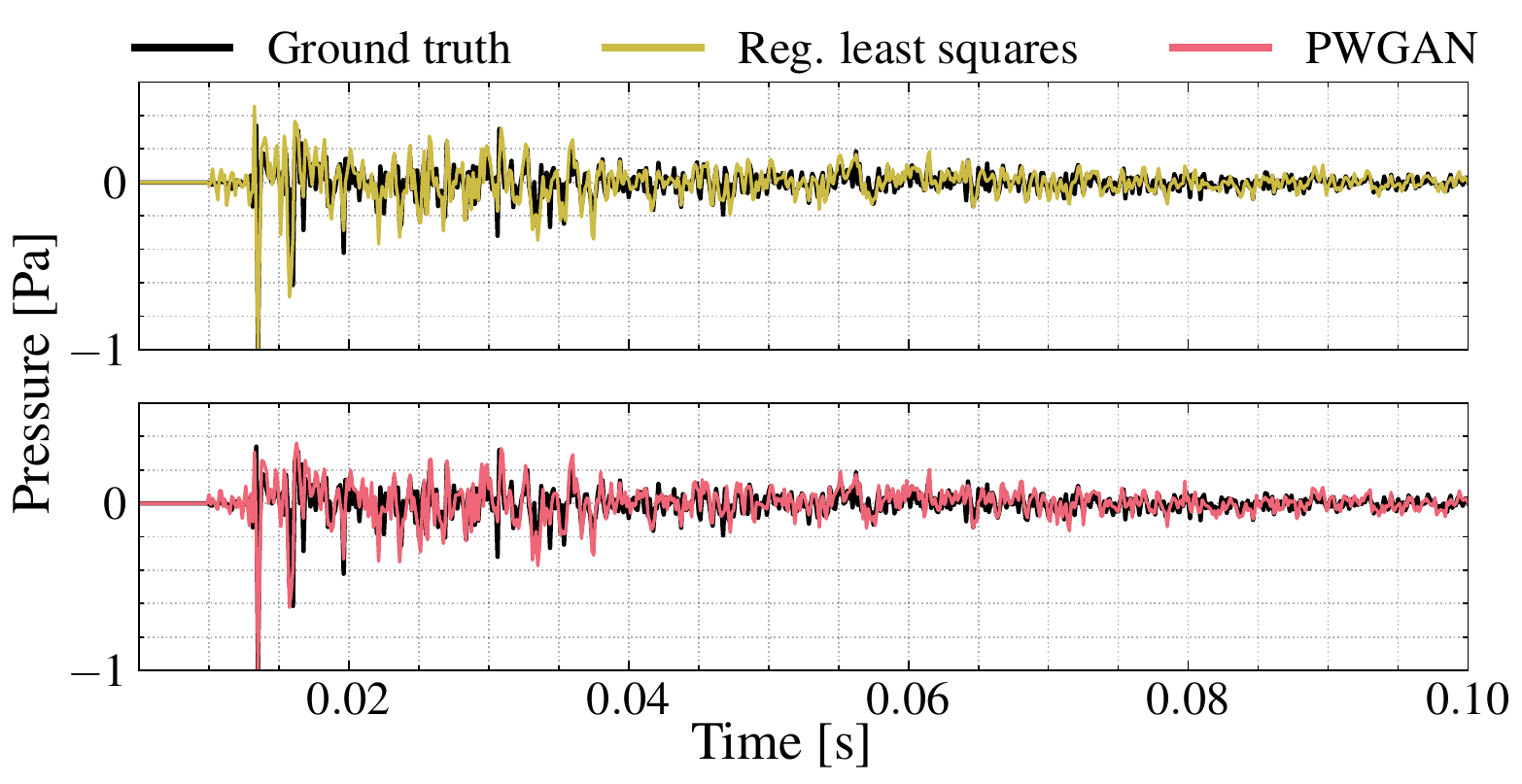}
{(a)}
\label{fig:interpRIR}
\end{subfigure}
\hspace{-0.04\linewidth} 
\begin{subfigure}
\centering
\includegraphics[trim={1mm 0 0mm 0mm},clip, width=\linewidth]{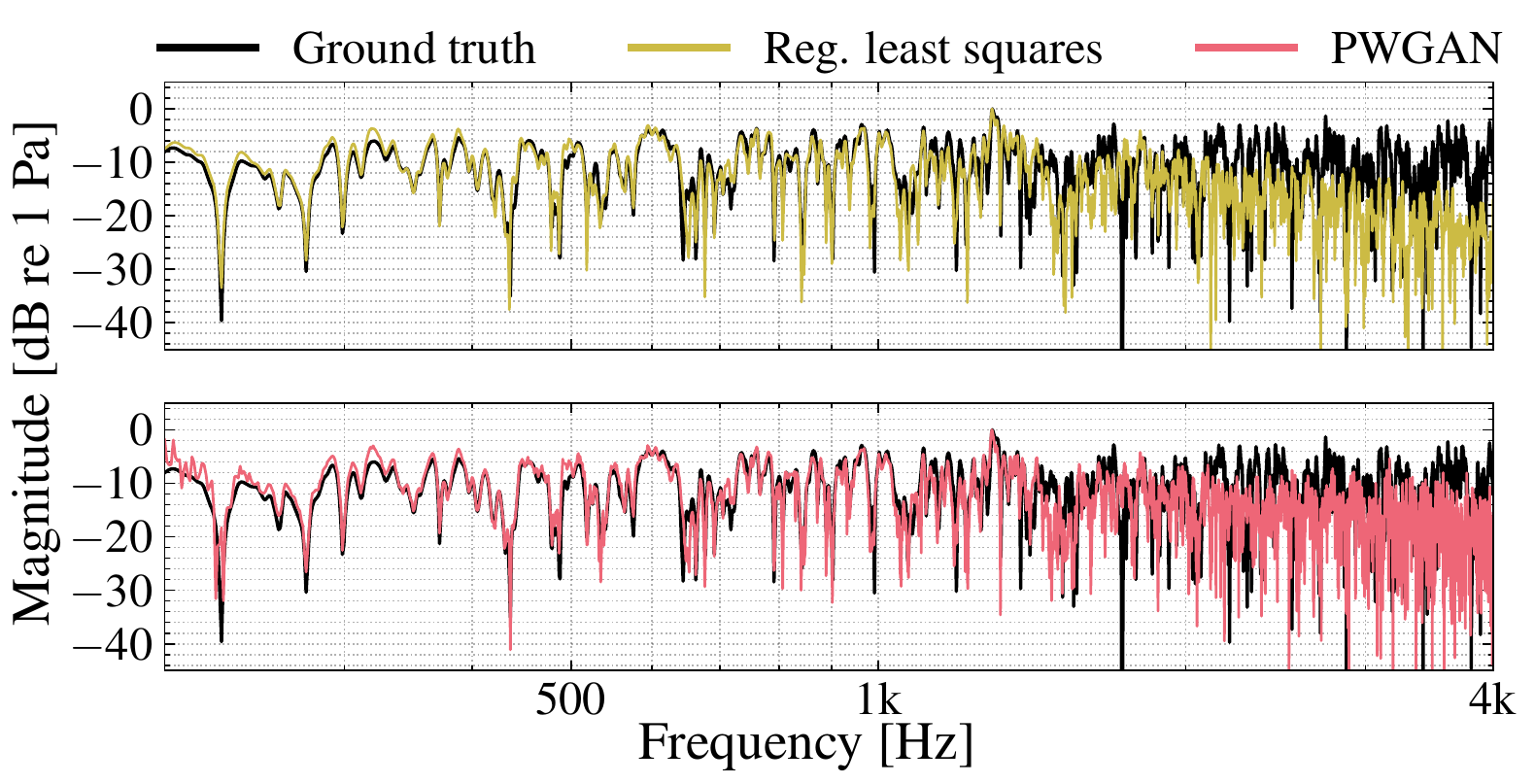}
{(b)}
\label{fig:interpFRF}
\end{subfigure}
\caption{\captionseven}
\label{fig:InterpRIRFRFcombined}
\end{figure}
\fi

In the extrapolated RIR of Fig. 8 at $\lvert r \rvert = 0.51$ m from the centre of the spherical array, we see a larger difference in terms of correlation, where the PWGAN achieves a correlation coefficient of 0.78 as opposed to 0.73 of the reg. least squares method. This becomes apparent when examining the amount of energy that each method tends to apply to the early reflections (before 0.03 s), with the Reg. least squares method overestimating these reflections as opposed to the PWGAN. Similar to the interpolated RIRs, the pre-ringing is reduced for the PWGAN reconstruction when compared to its Reg. least squares counterpart. Again, the extrapolated FRFs of Fig.\,\ref{fig:ExtrapRIRFRFcombined} reveal more energy in the higher frequency region but also more erroneous behaviour in the lower frequency region.
\newcommand{\captioneight}{(Color online) Extrapolated RIRs (a) and FRFs (b) for Reg. least squares (yellow) and PWGAN (red) with the respective correlation coefficients of 0.73 and 0.78 at $\lvert r \rvert$ = 0.51 m from the array centre.}
\ifdefined\showfigures
\begin{figure}[htb!]
\centering
\begin{subfigure}
\centering
\includegraphics[trim={0mm 1mm 1.5mm 0},clip, width=\linewidth]{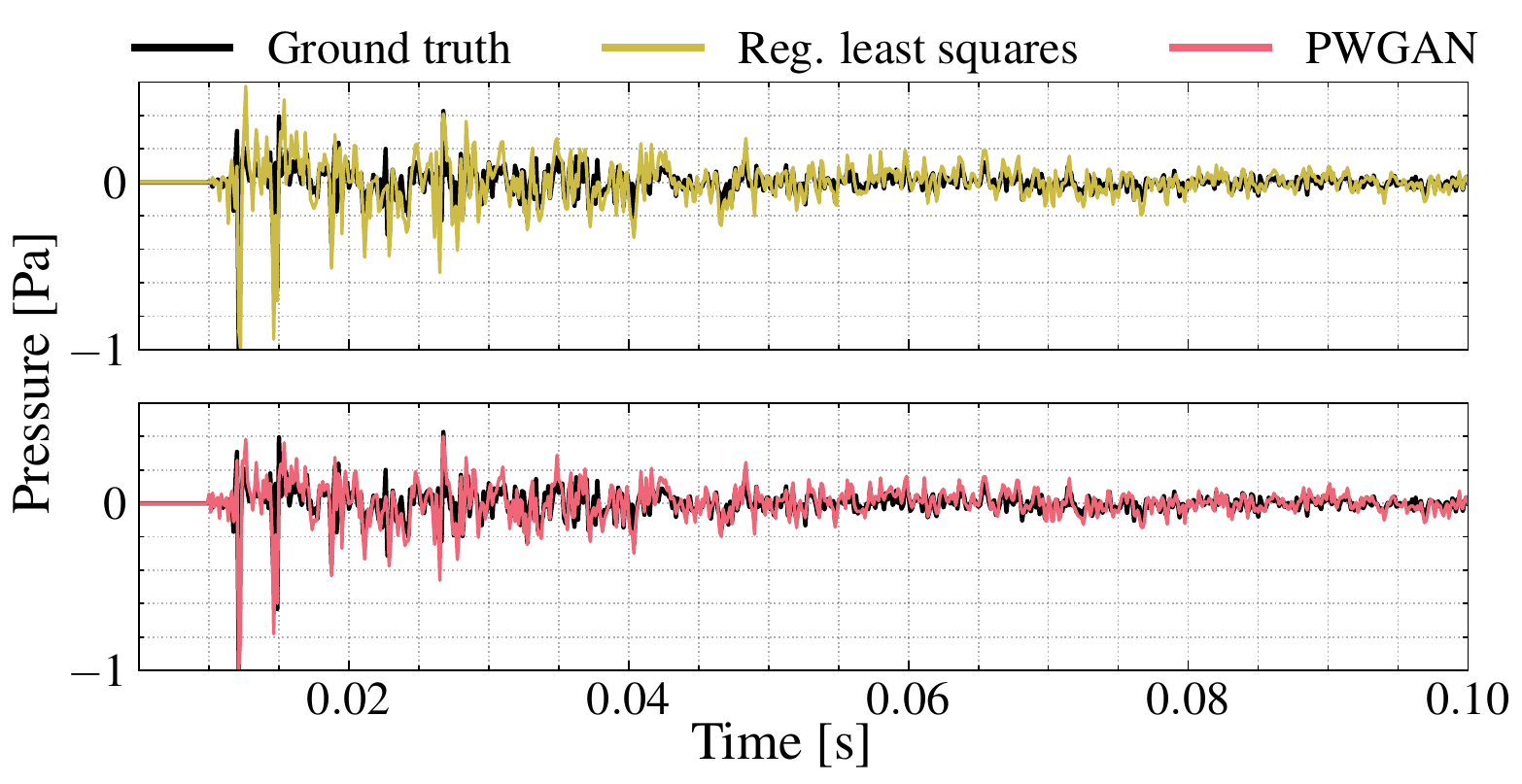}
{(a)}
\label{fig:extrapRIR}
\end{subfigure}
\hspace{-0.04\linewidth} 
\begin{subfigure}
\centering
\includegraphics[trim={1mm 0 0mm 0mm},clip, width=\linewidth]{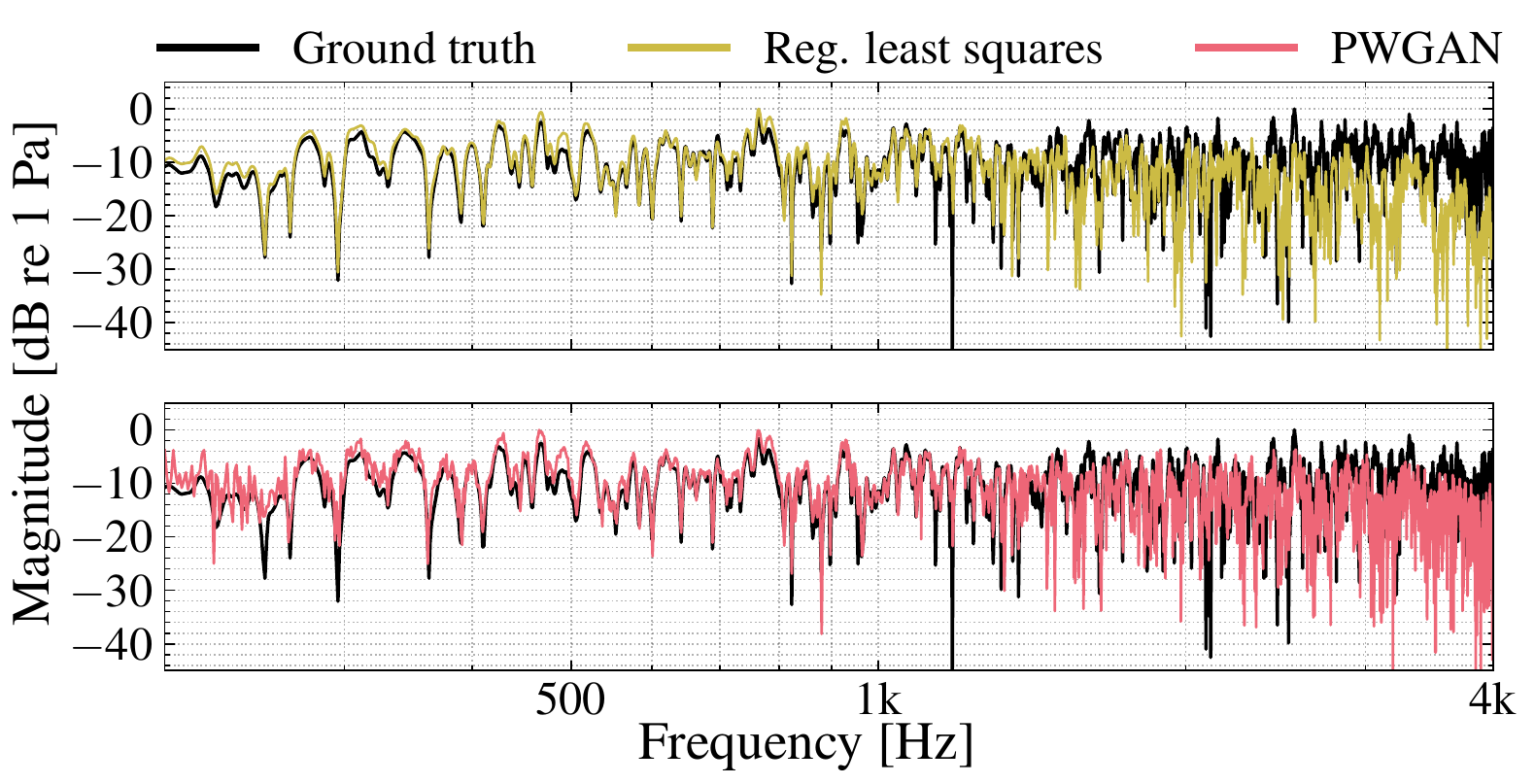}
{(b)}
\label{fig:extrapFRF}
\end{subfigure}
\caption{\captioneight}
\label{fig:ExtrapRIRFRFcombined}
\end{figure}
\fi

The spatial properties of the reconstructed sound fields are evaluated in Fig. 9 which shows both the SS and the NMSE of the reconstructions against the ground truth sound fields. Both graphs display the superior performance of the PWGAN in high frequencies, as the NMSE is lower for the PWGAN, while the SS is higher in the other, especially above 1 kHz. However, the network performance degrades at low frequencies below 400 Hz, where the reg. least squares method outperforms it. It is also evident that the network is largely inaccurate at certain few frequency bins starting at 1.1 kHz as is evident from the NMSE graph, but this is not apparent in the SS graph, suggesting that it could be due to a difference in scale since the SS metric is scale invariant.
\newcommand{\captionnine}{(Color online) Spatial Similarity (top) and NMSE (bottom) of the reconstruction plane of the DTU dataset for both the PWGAN and the Reg. least squares method.}
\ifdefined\showfigures
\begin{figure}[!ht]
    \centering
    \includegraphics[trim={0mm 0 0mm 0},clip, width = \linewidth]{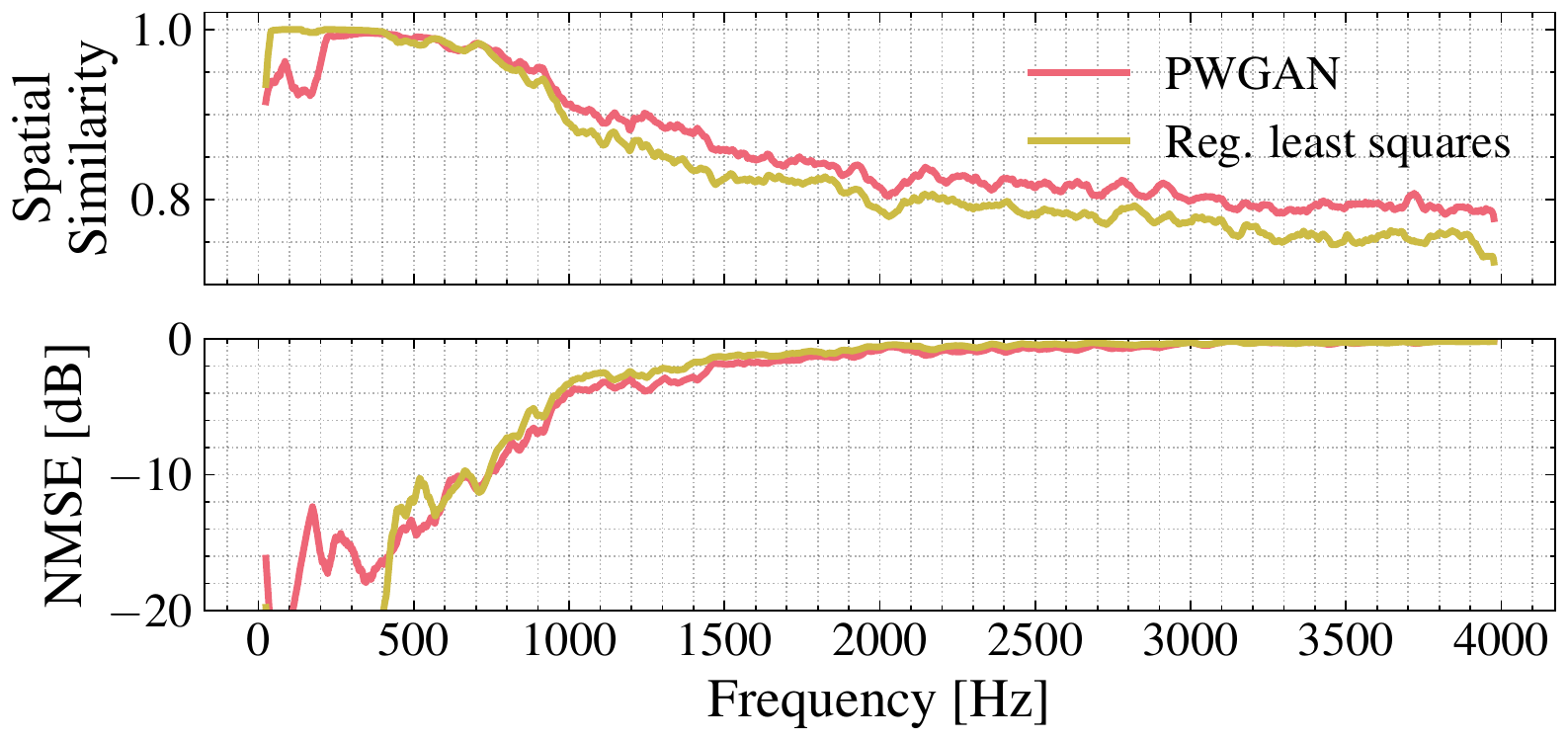}
    \caption{\captionnine}
    \label{fig:nmse_ss_dtu}
\end{figure}
\fi
\subsection{Planar sound field reconstruction}\label{subsec:planar_sf_reconstruction}
\noindent
The MeshRIR dataset sound fields are reconstructed using the PWGAN, a U-net (denoted as AE) and using the latent space optimisation of a Variational Auto-Encoder (VAE). Only the magnitudes are retained for this comparison as the AE can only reconstruct those due to limitations in its architecture which resemble that of image inpainting, with strictly positive activation functions. Fig. 10 shows each networks' performance as a function of frequency, both in terms of NMSE and SS. The top row of the aforementioned figure displays the measurement positions for 66 microphones superimposed on the complete set of measurements of various pure-tone sound field magnitudes for reference. For most frequencies examined, the PWGAN demonstrates both lower levels of NMSE and higher levels of SS, with the exception of frequencies below approximately 175 Hz and between 800 and 1000 Hz. However, it seems most networks seem to struggle in the lower frequency range, suggesting that the random wave model does not adequately describe the statistics of the sound field at low frequencies – as expected, in a bandwidth where room modes dominate. However, this poor performance may also be due to rank deficiency and the highly correlated columns of the dictionary, leading to a severely ill-conditioned problem at low frequencies. Conversely, the AE model demonstrates erratic behaviour at high frequencies above approximately 1200 Hz, in a range where the PWGAN is superior. In some cases even the VAE surpasses the AE in terms of performance for both metrics, for example, in terms of NMSE between 200 and 400 Hz and in terms of SS at 400 Hz.
\newcommand{\captionten}{(Color online) Planar sound field magnitude reconstruction using a fixed number (66) and position of microphones, evaluated in terms of NMSE (bottom left) and SS (bottom right) as functions of frequency for the PWGAN, AE and VAE neural networks. The top row displays the fixed measurement positions overlayed on the various pure-tone sound field magnitudes which were reconstructed.}
\ifdefined\showfigures
\begin{figure*}[htb!]
    \centering
    \includegraphics[trim=0.02in 0.02in 0.03in 0.in, clip,width=7in]{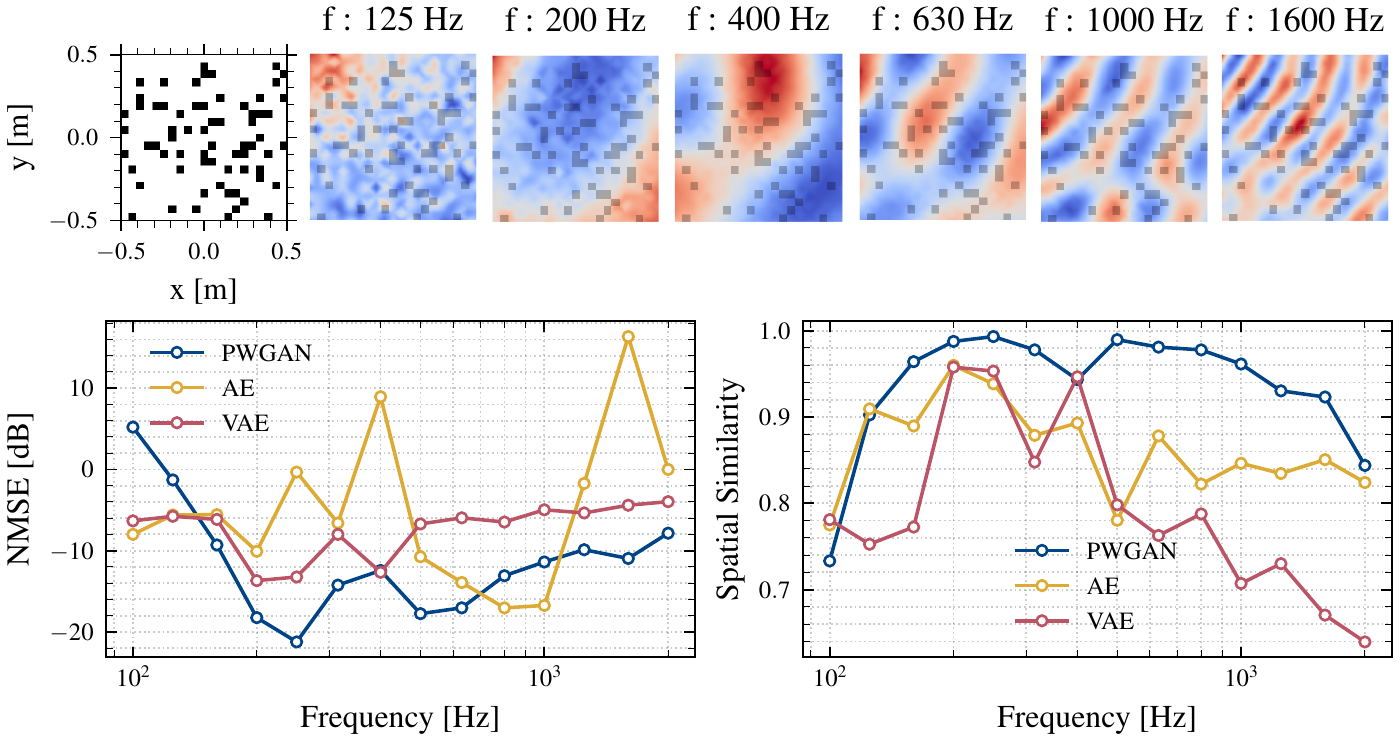}
    \caption{\captionten} 
    \label{fig:MeshRIR_error_vs_freq}
\end{figure*}
\fi

Figure 11 shows each networks' reconstruction performance for $f = 550$ Hz as a function of the quantity of measurement positions randomly distributed in space, both in terms of NMSE and SS. The top row of the aforementioned figure displays the increase in the number of measurement positions superimposed on the complete set of measurements for the $f = 550$ Hz sound field magnitude. In this case, the PWGAN constantly surpasses the baseline methods for both metrics. The VAE shows the worst performance for most of the measurement configurations.  However, it should be mentioned that the representation capabilities of the VAE are often hindered due to the limited size of the latent space, given that each factor of variation in the data space $\Vec{p}$ is represented by one coordinate in the latent space $\Vec{z}$ with only few degrees of freedom to represent a multi-modal data distributions (see \ref{Appendix:VAE_AE_details} for details). Regardless, the AE starts to fail above 220 microphone positions due to the fact that it has not been trained with more than 210 microphones, a problem that the other two methods do not seem to suffer from.
\newcommand{\captioneleven}{(Color online) Planar sound field magnitude reconstruction for a fixed number frequency $f = 550$ Hz evaluated in terms of NMSE (bottom left) and SS (bottom right) as functions of the quantity of microphones for the PWGAN, AE and VAE neural networks. The top row displays the sound field magnitude and the the various examples of measurement positions.}
\ifdefined\showfigures
\begin{figure*}[ht!]
    \centering
    \includegraphics[trim=0.02in 0.02in 0.03in 0in, clip,width=7in]{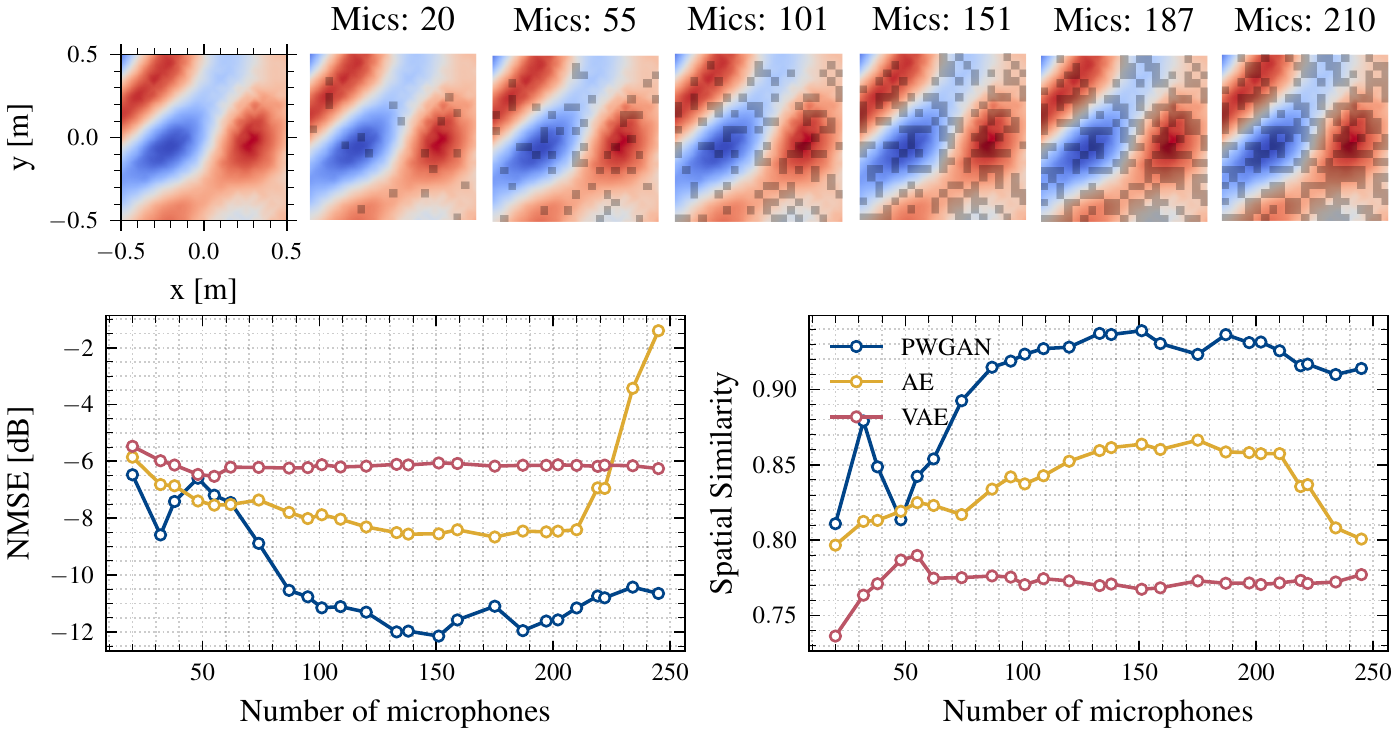}
    \caption{\captioneleven}
    \label{fig:MeshRIR_error_vs_nmics}
\end{figure*}
\fi
%


\section{Discussion}
\noindent
Overall, the results demonstrate the effectiveness of the PWGAN in reconstructing sound fields from microphone arrays or spatially distributed measurements (spherical and planar configurations were tested). The proposed PWGAN method was able to reconstruct room impulse responses (RIRs) more accurately than a classical plane wave expansion, and outperformed two state-of-the-art neural networks for sound field magnitude prediction.

One key finding is that the PWGAN is robust to out-of-distribution data, such as when the distribution of microphones has not been seen during the training phase. This can be attributed to the combination of using the GAN as well as a physically informed model, in this case a plane wave basis for decomposing the sound field. This also explains the fair performance when the number of measurement positions is limited. 

The PWGAN's ability to retain more energy at high frequencies and accurately reconstruct RIRs in extrapolated scenarios make it a valuable tool for a variety of applications in audio signal processing and acoustics. Whilst we acknowledge that the PWGAN's aptitude within the lower frequency range is comparatively weaker than in other frequency ranges, its efficacy and potential for sound field reconstruction applications is highly desirable, especially given its performance in the high frequency range. Regardless, this low frequency deficit seems to stem from the training data, the random wave model, which cannot accurately describe a sound field with a significantly low modal overlap, in a frequency range below the Schroeder frequency of a room.\cite{fjacobsen}
\section{Conclusions}
\noindent
In this paper, we proposed a method for sound field reconstruction using Generative Adversarial Networks to accurately reconstruct sound fields from a limited set of measurements. The GAN was trained using a random wave field model, which was generated in real time during the adversarial process. This allowed for efficient data-free training with physically sensible priors to learn the underlying statistical distributions of pressure in rooms. We demonstrated the effectiveness of our method by reconstructing room impulse responses in a lightly damped room using a spherical microphone array, as well as the sound field magnitudes for irregular microphone distributions and at various frequencies. At the same time, we compare the proposed method with state-of-the-art methods for sound field reconstruction and show that our method improves the estimated pressure fields both in terms of accuracy and energy retention at high frequencies, where classical methods exhibit a comparatively weaker performance. 

Looking ahead, generative models offer promising avenues for improving sound field estimation and data generation in acoustics. The ability to learn statistical distributions from limited measurements has the potential to bring advances when used in acoustics, allowing for more efficient and accurate modelling of sound propagation in complex environments. Further research could explore the use of generative models for tasks such as source localisation and separation, as well as investigating their potential for use in real-time applications. Overall, our work highlights the potential of generative models in acoustics and provides a framework for further exploration and development of these techniques.
\begin{acknowledgments}
\noindent
This work was supported by the VILLUM Foundation, under grant number 19179, ``Large-scale acoustic holography.'' The authors would like to thank Antonio Andr\'es Figeroa Dur\'an, Diego Caviedes Nozal, Franz Heuchel and Manuel Hahmann for the valuable discussions during the conception of this work.
\end{acknowledgments}

\appendix*
\section{Baseline training \& architecture}\label{Appendix}
\subsection{Partial Convolutions}
\noindent
Unlike standard convolutions, partial convolutions offer the flexibility to handle irregularly distributed spatial data, making them suitable for tasks such as inpainting or processing inputs with missing information. By using a sliding convolutional window and incorporating a mask that selectively weights the input, partial convolutions adjust the contribution of each sample point in the convolutional operation, allowing for adaptive and context-aware computations. The resulting output feature maps are computed by considering both the input features and their corresponding mask, resulting in effective spatial feature extraction while accommodating missing or incomplete data.
\subsection{Sound field auto-encoder and variational auto-encoder training and architectures}\label{Appendix:VAE_AE_details}
The training process and architecture of both neural networks of sections \ref{subsec:sf_prediction} and \ref{subsec:planar_sf_reconstruction} are described hereon. The Auto-Encoder is adapted from Llu{\'\i}s et al.\citep{sfrecon_inpainting} to accommodate sound field magnitudes of dimensionality, $(21, 21, 1)$ and the architecture used can be seen in Tab. \ref{tab:ae_architecture}. It was trained for 50000 iterations with a batch size of 32 with a learning rate of $\eta_{\text{ae}} = 0.0004$ and the Adam optimiser, by minimising the mean-square error between observations and predictions. This process took approximately 21 hours on a single GPU.
\begin{table}[!t]
\renewcommand{\arraystretch}{1.3}
\caption{Summary of Auto-Encoder (U-net) network}
\label{tab:ae_architecture}
\resizebox{\reprintcolumnwidth}{!}{%
\begin{tabular}{lll}
\textbf{Layer Name} & \textbf{Input Shape} & \textbf{Description} \\
\hline\hline
\texttt{input sound field} & (21, 21, 1) & Batch input layer \\
\texttt{input mask} & (21, 21, 1) & Batch input layer \\
\texttt{encoder partial conv. 1} & [(21, 21, 1), (21, 21, 1)] & Partial convolutional layer (64 filters, kernel size: (5, 5)) \\
\texttt{activation 1} & (11, 11, 64) & ReLU activation \\
\texttt{encoder partial conv. 2} & [(11, 11, 64), (11, 11, 64)] & Partial convolutional layer (128 filters, kernel size: (3, 3)) \\
\texttt{encoder BN 1} & (6, 6, 128) & Batch normalization \\
\texttt{activation 2} & (6, 6, 128) & ReLU activation\\
\texttt{encoder partial conv. 3} & [(6, 6, 128), (6, 6, 128)] & Partial convolutional layer (256 filters, kernel size: (3, 3)) \\
\texttt{encoder BN 2} & (3, 3, 256) & Batch normalization \\
\texttt{activation 3} & (3, 3, 256) & ReLU activation  \\
\texttt{encoder partial conv. 4} & [(3, 3, 256), (3, 3, 256)] & Partial convolutional layer (512 filters, kernel size: (3, 3)) \\
\texttt{encoder BN 3} & (3, 3, 512) & Batch normalization \\
\texttt{activation 4} & (3, 3, 512) & ReLU activation  \\
\texttt{concatenate 1} & [(3, 3, 512), (3, 3, 256)] & Concatenate activation 4 and activation 3\\
\texttt{decoder partial conv. 1} & [(3, 3, 768), (3, 3, 768)] & Partial convolutional layer (256 filters, kernel size: (3, 3))\\
\texttt{decoder BN 1} & (3, 3, 256) & Batch normalization \\
\texttt{activation 5} & (3, 3, 256) & Leaky ReLU activation \\
\texttt{decoder upsampling 1} & (3, 3, 256) & Nearest neighbor upsampling \\
\texttt{concatenate 2} & [(6, 6, 256), (6, 6, 128)] & Concatenate decoder upsampling 1 and activation 2\\
\texttt{decoder partial conv. 2} & [(6, 6, 384), (6, 6, 384)] & Partial convolutional layer (128 filters, kernel size: (3,3))\\
\texttt{decoder BN 2} & (6, 6, 128) & Batch normalization \\
\texttt{activation 6} & (6, 6, 128) & Leaky ReLU activation \\
\texttt{decoder upsampling 2} & (6, 6, 128) & Nearest neighbor upsampling \\
\texttt{decoder crop 1} & (12, 12, 128) & Crop dimensions of feature space\\
\texttt{concatenate 3} & [(11, 11, 128), (11, 11, 64)] & Concatenate decoder crop 1 and activation 1\\
\texttt{decoder partial conv. 3} & [(11, 11, 192), (11, 11, 192)] & Partial convolutional layer (64 filters, kernel size: (3,3))\\
\texttt{decoder BN 3} & (11, 11, 64) & Batch normalization \\
\texttt{activation 7} & (11, 11, 64) & Leaky ReLU activation \\
\texttt{decoder upsampling 3} & (11, 11, 64) & Nearest neighbor upsampling \\
\texttt{decoder crop 2} & (22, 22, 64) & Crop dimensions of feature space\\
\texttt{concatenate 4} & [(21, 21, 1), (21, 21, 64)] & Concatenate decoder crop 2 and input\\
\texttt{decoder partial conv. 4} & [(21, 21, 65), (21, 21, 65)] & Partial convolutional layer (1 filter, kernel size: (5,5))\\
\texttt{activation 8} & (21, 21, 1) & Leaky ReLU activation\\
\texttt{output conv.} & (21, 21, 1) & output 2D convolution layer (1 filter, kernel size: (5,5))\\
\texttt{activation 9} & (21, 21, 1) & Sigmoid activation \\
\end{tabular}
}
\end{table}

The Variational Auto-Encoder (VAE) architecture can be seen in Tab. \ref{tab:vae_architecture}. It was trained using the Adam optimiser with a learning rate of $\eta_{\text{vae}} = 0.0001$ and a batch size of 32 for 40000 iterations. The training objective was the Kullback-Leibler (KL) divergence and the reconstruction likelihood.\citep{vae} The KL divergence term measures the dissimilarity between the learned latent distribution and a predefined prior distribution, encouraging the VAE to generate diverse and representative latent codes. The reconstruction likelihood term quantifies the quality of the reconstructed output compared to the original input in terms of the mean-squared error. The VAE train approximately 20 hours for the aforementioned amount of iterations on a single GPU. The same training data was used for all three neural networks (e.g. random wave fields).

\begin{table}[h]
\renewcommand{\arraystretch}{1.3}
\centering
\caption{Summary of Variational Auto-Encoder network}
\label{tab:vae_architecture}
\resizebox{\reprintcolumnwidth}{!}{%
\begin{tabular}{lll}
\textbf{Layer Name} & \textbf{Input Shape} & \textbf{Description} \\
\hline\hline
\texttt{conv2d 1} & (21, 21, 2) & Convolutional layer (32 filters, kernel size: (3, 3), stride: (1, 1))\\
\texttt{activation 1} & (10, 10, 32) & ReLU activation \\
\texttt{conv2d 2} & (10, 10, 32) & Convolutional layer (64 filters, kernel size: (3, 3), stride: (1, 1)) \\
\texttt{activation 2} & (4, 4, 64) & ReLU activation \\
\texttt{flatten} & (4, 4, 64) & Flatten layer \\
$\Vec{z}_\sigma$ & (1024) & Dense layer (16 units, linear activation) \\
$\Vec{z}_\mu$ & (1024) & Dense layer (16 units, linear activation) \\
$\mathcal{N}(\Vec{z}_\mu, \Vec{z}_\sigma)$ & (16) & Lambda layer (sampling from normal distribution) \\
\texttt{dense 1} & (16) & Fully connected layer (8192 units, linear activation) \\
\texttt{reshape} & (8192) & Reshape layer \\
\texttt{conv2d transp. 1} & (16, 16, 32) & Convolutional transpose layer (64 filters, kernel size: (3, 3), stride: (2, 2)) \\
\texttt{activation 3} & (32, 32, 64) & ReLU activation \\
\texttt{conv2d transp. 2} & (32, 32, 64) & Convolutional transpose layer (32 filters, kernel size: (3, 3), stride: (2, 2)) \\
\texttt{activation 4} & (64, 64, 32) & ReLU activation \\
\texttt{conv2d transp. 3} & (64, 64, 32) & Convolutional transpose layer (2 filters, kernel size: (3, 3), stride: (1, 1), activation: Linear) \\
\texttt{reshape} & (64, 64, 2) & Reshape layer to (4092,2)\\
\end{tabular}%
}
\end{table}
\bibliography{biblio.bib}
\ifdefined\appendcaptions
\section*{Figure captions}
\renewcommand{\labelenumi}{Figure \arabic{enumi}:}
\begin{enumerate}
\setlength{\itemindent}{5ex}
\item{\captionone}
\item{\captiontwo}
\item{\captionthree}
\item{\captionfour}
\item{\captionfive}
\item{\captionsix}
\item{\captionseven}
\item{\captioneight}
\item{\captionnine}
\item{\captionten}
\item{\captioneleven}
\end{enumerate}
\fi
\end{document}